%% file: Paper.tex
\documentclass[letterpaper,twocolumn,10pt]{article}
\usepackage{usenix}

\usepackage{enumitem}
\usepackage[table]{xcolor}
\usepackage{xurl}
\usepackage{booktabs}
\usepackage{multirow}
\usepackage{multicol}
\usepackage{rotating}
\usepackage{makecell}
\usepackage{amssymb}
\usepackage{pifont}
\usepackage{listings}
\usepackage{amsmath}
\usepackage{tabularx}
\usepackage{caption}
\usepackage{tikz}

\usetikzlibrary{decorations.pathreplacing,calligraphy} %

\newcommand{\name}{\textsc{CATS}}
\newcommand{\datasetname}{ERPCorp}

\newcommand{\grayrow}{\rowcolor{gray!10}}

\newenvironment{tightemize}{%
    \begin{itemize}[leftmargin=*]%
    \setlength{\itemsep}{0pt}%
    \setlength{\parskip}{0pt}%
    \setlength{\parsep}{0pt}
}{%
    \end{itemize}
}

  {\begin{enumerate}[leftmargin=*]%
            \setlength{\itemsep}{0pt}%
            \setlength{\parskip}{0pt}%
            \setlength{\parsep}{0pt}}%
     {\end{enumerate}}

\begin{document}

\date{}

\title{\Large \bf Can Risk-Based Alerting Mitigate Cybersecurity Alert Fatigue?}

\author{
    {\rm Rafael Uetz}\\
    Fraunhofer FKIE
    \and
    {\rm Philipp Bönninghausen}\\
    Fraunhofer FKIE
    \and
    {\rm Louis Hackländer-Jansen}\\
    Fraunhofer FKIE
    \and
    {\rm Martin Henze}\\
    RWTH Aachen University \& Fraunhofer FKIE
}

\maketitle

\begin{abstract}
    Security operations centers (SOCs) face large numbers of false alerts, making detection of cyberattacks difficult under typical resource constraints.
    Risk-based alerting (RBA) has been proposed as a means to reduce false alerts and has reportedly succeeded in doing so in various enterprise deployments.
    However, RBA has not been comprehensively evaluated until now, leaving implementation mostly guesswork based on anecdotal evidence.
    In this paper, we present the first systematic evaluation of RBA.
    To this end, we reformulate it as a continuous alert prioritization problem rather than a binary decision problem (i.e., whether an alerting threshold is exceeded), allowing us to evaluate performance across all possible thresholds and thus model SOCs of varying sizes and alert volumes.
    We distill five fundamental risk hypotheses, formalize them as independently parametrizable modules, and implement them in our novel experimentation suite CATS.
    We thoroughly assess the hypotheses across eight diverse alert datasets, six of which we created or extended to make such an evaluation possible.
    Our results show that certain combinations of hypotheses achieve a remarkable alert prioritization performance (AUROC $\mu=0.92$, $\sigma=0.09$ across the eight datasets), outperforming a straightforward prioritization by alert severity level (AUROC $\mu=0.72$, $\sigma=0.21$).
    We conclude that RBA can substantially reduce the number of false alerts that analysts have to review and thus has the potential to mitigate cybersecurity alert fatigue.
    In addition, it serves as a strong baseline for more complex, resource-intensive alert triage approaches (e.g., based on large language models).
\end{abstract}

\section{Introduction}
\label{sec:introduction}

Large numbers of organizations fall victim to successful cyberattacks~\cite{verizon2026dbir}.
Risk managers even deem cyber incidents to be the most important global business risk, according to a recent survey~\cite{allianz2026risk}.
To mitigate this risk, organizations need to invest in both prevention \emph{and} detection of cyberattacks, since preventive measures alone are insufficient~\cite{lord2022investments}.

Consequently, many organizations establish a dedicated (internal or external) \emph{security operations center} (SOC), where security experts (\emph{analysts}) continuously review potential indicators of security incidents (i.e., \emph{cybersecurity alerts}) from a variety of sources throughout the enterprise network, trying to discover attackers and ultimately stop them before they can reach their final goals~\cite{vielberth2020soc}.
However, both the size and diversity of modern enterprise networks with respect to users, hosts, applications, and security systems usually lead to an excessive number of alerts that must be reviewed, where only a small fraction are actual indicators of an attack and the rest are false alerts~\cite{alahmadi2022false, axelsson2000base}.
Analysts may thus be overwhelmed by false alerts and therefore tend to miss actual attacks, a phenomenon known as \emph{(cybersecurity) alert fatigue}~\cite{shahroz2025alertfatigue, sans2024detectionsurvey,nobles2022stress}.

To mitigate these issues, both researchers and practitioners have been working on ways to reduce the number of false alerts that SOC analysts need to review (cf. Sections~\ref{sec:background} and~\ref{sec:relatedwork}).
While in an ideal world, each security system could be tuned to emit true alerts only, this is often not possible in practice due to insufficient  configurability, analyst expertise, or discriminability of benign and adversarial behavior (which can only be decided in the context of additional information)~\cite{alahmadi2022false}.
For this reason, a multitude of approaches exist to cope with highly noisy alerts and provide analysts with fewer, higher-quality alerts by means of automated alert prioritization~\cite{jalalvand2024alert}, aggregation~\cite{landauer2022dealing}, or correlation~\cite{navarro2018multistep}.

Besides primarily academic approaches, which often employ machine learning~(cf. Section~\ref{sec:relatedwork}), one contrasting approach called \emph{Risk-Based Alerting} (RBA) stands out as it has been widely adopted by practitioners, especially in the Splunk~\cite{splunk2026splunk} community.
RBA is fully explainable, requires no training data, and is substantially less computationally expensive than large language model-based methods~\cite{navarro2018multistep}.
Its fundamental idea is to estimate the cybersecurity \emph{risk} for entities such as users or hosts by weighting entity-related alerts based on so-called \emph{risk (incident) rules} that capture broadly valid knowledge on security monitoring and cyberattacks, e.g., ``multiple different rules triggering in temporal proximity for the same entity indicate a high risk''.

Manual review is only triggered when an entity's risk exceeds a certain threshold.
For example, while a crashing PDF reader, a newly created user, and a high outbound traffic might not justify individual alerts for manual review, their occurrence on the same system within one day might indicate malware infection and consequential data exfiltration.

Despite its apparent success in practice, to our knowledge RBA has not been systematically evaluated until now.
This is probably due to (1)~a lack of publicly available alert datasets that capture a broad range of realistic environments, benign activity, and cyberattacks~\cite{boenninghausen2024comidds} and (2)~a lack of formalization of the RBA methodology along with suitable metrics to measure potential improvements for analysts.
Without a systematic evaluation, it remains unknown whether the anecdotally reported success generalizes beyond individual deployments, leaving SOCs that adopt RBA at risk of fruitless effort or even an increase in missed true alerts.

In this paper, we address these issues to answer not only the question \emph{if} RBA can mitigate cybersecurity alert fatigue, but also \emph{how} it should be implemented and configured to achieve good results.
To this end, we review literature on RBA and systematize the suggested rules by \emph{explicitly formulating their underlying risk hypotheses} (i.e., assumptions on how to distinguish benign from adversarial activity based solely on alerts).
We define five risk hypotheses suitable for evaluation.
They address (1)~the severity level of triggered detection rules as well as the spatio-temporal (2)~accumulation, (3)~variety, (4)~rarity, and (5)~periodicity of alerts with respect to their affected entity and triggered detection rule.

We present a comprehensive experimentation suite called \name{} for evaluating and visually exploring risk hypotheses using different alert datasets and metrics.
We implemented each of the five risk hypotheses as a configurable \emph{risk module} that assigns a \emph{risk score} to each alert based on the underlying hypothesis and a set of relevant alerts within the dataset (e.g., alerts occurring within one hour on the same host).
Risk modules can be combined into \emph{pipelines}.

We evaluate the five selected risk hypotheses using CATS against eight alert datasets that contain multiple alert sources (Falco, Sigma, Suricata, and Wazuh).
These alert datasets stem from five diverse environments (namely, one enterprise network and four cybersecurity testbeds) and are thus based on heterogeneous cyberattacks and benign activity.
Our results show that a weighted combination of the five risk modules achieves a remarkable prioritization performance across all datasets (AUROC $\mu=0.92$, $\sigma=0.09$), which is a massive improvement over no prioritization ($\mu=0.5$, $\sigma=0$), substantially better than a straightforward prioritization by rule severity level ($\mu=0.72$, $\sigma=0.21$), and still considerably ahead of the best single risk module ($\mu=0.87$, $\sigma=0.17$).

Overall, our work indicates that RBA -- if implemented and configured as worked out in this paper -- can substantially increase the efficiency and effectiveness of cyberattack detection and may thus decrease cybersecurity alert fatigue, with almost negligible computational cost.
Researchers and practitioners can use \name{} to reproduce and tune our results, optionally using their own alert datasets.

In summary, we make the following contributions:
\begin{tightemize}
    \item We reformulate RBA to perform continuous alert prioritization rather than fixed-threshold alerting, thereby enabling a more comprehensive evaluation and allowing analysts to review alerts more efficiently in order of priority (§\ref{sec:problem}).
    \item We analyze literature on RBA and identify five fundamental hypotheses to distinguish benign activity from cyberattacks solely using basic alert characteristics such as timestamp, triggered detection rule, and source host (§\ref{sec:hypotheses}).
    \item We present the open-source tool CATS for visual exploration and automated evaluation of alert prioritization methods such as the five identified risk hypotheses (§\ref{sec:cats}).
    \item We contribute new, labeled alert datasets based on the log dataset DEDALE, the testbed SOCBED, and the MITRE APT29 emulation plan ``Scenario 2'' (§\ref{sec:datasets}).
    \item We evaluate the five risk hypotheses against the above datasets as well as an enterprise and a preexisting dataset (AIT-ADS) using CATS, showing that they can substantially reduce false alerts, especially when combined (§\ref{sec:evaluation}).
\end{tightemize}

\noindent\textbf{Open Science Statement.}
We make our artifacts publicly available (see the Open Science Appendix for details).
In particular, an anonymized version of CATS can be tested at \href{https://962012d09b.github.io/cats_webapp/}{https://962012d09b.github.io/cats\_webapp/}.

\section{Background: Risk-Based Alerting}
\label{sec:background}

To lay the foundation for this work, we give a brief overview of cybersecurity alert sources, alert triage, and risk-based alerting in enterprise networks.

\textbf{Alert Sources and Triage.}
The core functions of a SOC are alert triage, threat detection, and incident response~\cite{sans2025socsurvey}.
Alert triage denotes the process of deciding whether an alert is justified (and should thus be further investigated) or a false positive~\cite{vielberth2020soc}.
On the technical side, alert triage is usually performed by using a central Security Information and Event Management (SIEM) system, which collects log data and alerts from various sources~\cite{bhatt2014siem}.
Alert sources can be host- or network-based, e.g., endpoint security systems, intrusion detection systems, and the SIEM system itself~\cite{sans2024socsurvey}.

Aside from commercial products (e.g., Microsoft Defender Antivirus~\cite{microsoftdefender}), organizations commonly rely on open-source systems for threat detection.
In particular, the evaluation in this work is based on alerts generated by four widespread open-source systems, namely, \emph{Falco}~\cite{falco} (a rule-based threat detection system primarily for Linux, Kubernetes, and cloud events), \emph{Sigma}~\cite{sigma} (a generic detection rule format for SIEM systems with a focus on Windows threats), \emph{Suricata}~\cite{suricata} (a high-performance, network-based analysis and threat detection software), and \emph{Wazuh}~\cite{wazuh} (an open-source Extended Detection \& Response and SIEM system).

\begin{figure}[t]
    \centering
    \includegraphics[width=\columnwidth]{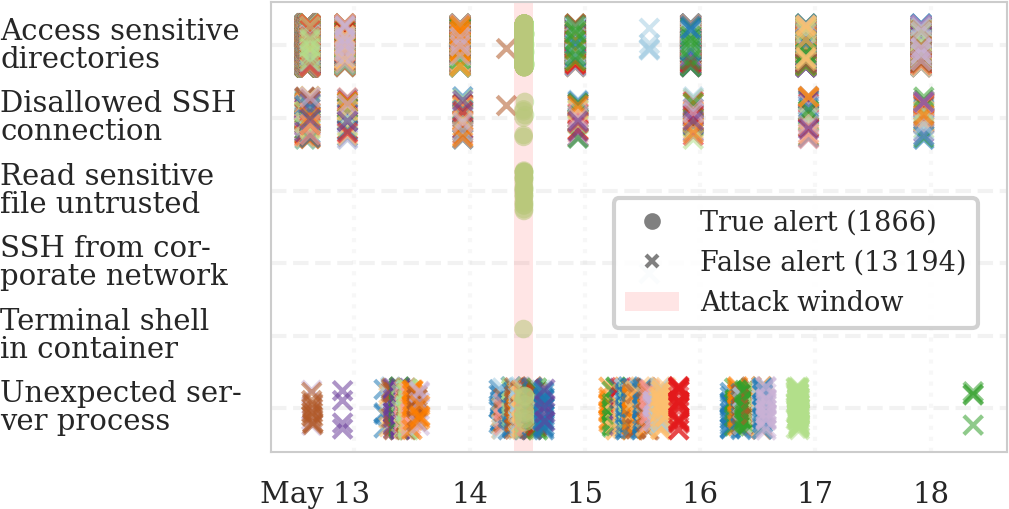}
    \caption{Visualization of our ``\datasetname{} Falco'' dataset, showing the six rules that triggered a total of 15\,060 alerts during the one-week collection period (colors represent hosts).}
    \label{fig:erpcorp}
\end{figure}

Each of these systems comes with regularly updated and fully customizable threat detection rules (consider the Sigma documentation for exemplary rules~\cite{sigmarules}).
To convey an impression of the multitude of alerts that SOCs are facing, Figure~\ref{fig:erpcorp} shows alerts triggered by Falco rules in an actual enterprise network over one week, including a simulated attack (cf. Section~\ref{sec:datasets}).
Despite Falco being just one of multiple security systems deployed in this enterprise, its number of alerts might already outpace analyst capacity.

\textbf{Risk-Based Alerting.}
RBA was first proposed in 2018 as a mitigation for having too many false alerts in a SOC~\cite{apger2018goodbye}.
While being popularized by Splunk and its community~\cite{splunk2024rba}, similar concepts have also been adopted in other SIEM systems such as Elastic~\cite{elastic2026entityanalytics}.
Its basic idea is as follows:
Instead of reviewing each alert from every threat detection system individually, all alerts are first centrally collected in a \emph{risk index} and assigned a \emph{risk score} by the use of expert-written \emph{risk rules}.
Such a  risk rule may, e.g., assign the risk score based on the original severity level of the triggered detection rule.
The score could then be adapted by a \emph{risk modifier} depending on the business criticality of the affected host or user.
For example, an alert from a domain controller would receive a higher risk score than from a regular system.

In this step, the alert is also normalized and potentially enriched such that attributes for the subsequent steps are available (e.g., timestamp, detection rule name, and affected entity).
Next, so-called \emph{risk incident rules} are applied to the risk index.
Their purpose is to correlate the original alerts and generate a risk-driven alert (called \emph{notable}) for manual review if and only if high-risk activity is found.
For example, a notable could be generated when three different detection rules trigger for a certain host or user within one day.
Note that RBA does not involve any learning phase, offsetting it from machine learning-based approaches. %

Several SOCs that adopted RBA report substantial improvements in detection efficiency, e.g., a true positive rate rising from 7.07\% to 33\%~\cite{apger2019modernize}, a false positive rate reduced from 78\% to 48\%~\cite{stearns2021accenture}, and an 80\%~\cite{bjerke2023eliminate} or even 800x~\cite{chila2021supercharge} reduction in the number of alerts.
As this anecdotal evidence is promising, we aim to perform a systematic and reproducible evaluation of RBA with the goal of providing concrete guidance to practitioners and foster further research.

\input{RIRs}

\section{Problem Statement, Goal, and Approach}
\label{sec:problem}

Our starting point is an organization with the goal of detecting adversarial activity within their network.
To this end, they operate threat detection systems that produce alerts on potential evidence of misuse.
Due to imperfect detection rules~\cite{sans2024detectionsurvey} paired with a low base rate of attacks~\cite{axelsson2000base}, alerts are mostly false~\cite{alahmadi2022false} and exceed analysts' resources for manual review, bearing the risk of missing true alerts~\cite{shahroz2025alertfatigue}.

Our high-level goal is enabling analysts to discover more true alerts given the same time resources or, equivalently, discovering the same number while reducing time spent (thus increasing SOC efficiency and likely reducing alert fatigue).
For this purpose, we evaluate the suitability of the RBA methodology as described above, but reformulate it to enable an expressive evaluation.
To this end, we combine the notion of risk rules (which assign risk scores to individual alerts) and risk incident rules (which correlate alerts with already-assigned risk scores) such that the latter also modify the risk scores of alerts instead of generating notables.
This turns RBA from a binary decision problem (notable generated or not) into a continuous  \textbf{prioritization} problem, where original alerts can be ranked by their risk scores.

This reformulation enables analysts to review alerts in descending order of risk score until their time resources are depleted, whereas manually-defined thresholds will usually generate either too many or too few notables.
In addition, it allows us to evaluate \emph{all} possible thresholds instead of just one by using established metrics such as AUROC (area under the receiver operating characteristic curve), AP (average precision), and Brier score (cf. Section~\ref{sec:metrics}), thereby modeling SOCs of varying sizes and alert volumes.

Deviating from the RBA nomenclature, we denote the combined rules as \textbf{risk hypotheses} to clearly emphasize that each rule represents a detection hypothesis with the goal of assigning a higher risk score to true alerts than to false alerts.
Note that, since risk hypotheses combine risk rules and risk incident rules, they can assign risk scores to an alert based on (1) information contained in the alert itself (e.g., its severity level), (2) information contained in other alerts (e.g., their historical frequency), or (3) both.
For example, one risk hypothesis could be: \emph{``A high-severity alert preceded by other alerts on the same host within a short time frame indicates a high risk''} (since benign activity is unlikely to trigger temporally close \emph{and} high-severity alerts).
We call a concrete implementation of a risk hypothesis a \textbf{risk module}.
Such a module comprises a formula for calculating a risk score for each alert as well as parameters to influence this calculation (e.g., the considered time frame).

Note that we deliberately refrain from evaluating entity risk modifiers (which adapt the risk score, e.g., for certain hosts or users) because modifier definition is largely subjective and thus potentially biased (especially when attacks in datasets are known).
Moreover, we do not assign risk scores to entities instead of alerts, as is the case for Splunk's RBA implementation~\cite{splunk2024rba}.
Doing so would leave the analyst with the task of investigating why an entity's risk is high, which is equivalent to triaging the contributing alerts and therefore not an advantage in our opinion.
Finally, we assume that analysts require the same time for each alert to triage.
While this might be an oversimplification, we are not aware of any practice-approved model for estimating triage time for a given alert.

To prepare the ground for our evaluation of RBA, we will now introduce candidate risk hypotheses (Section~\ref{sec:hypotheses}), implement them in our novel experimentation suite (Section~\ref{sec:cats}), and finally assemble evaluation datasets (Section~\ref{sec:datasets}).

\begin{figure*}[t]
    \centering
    \includegraphics[]{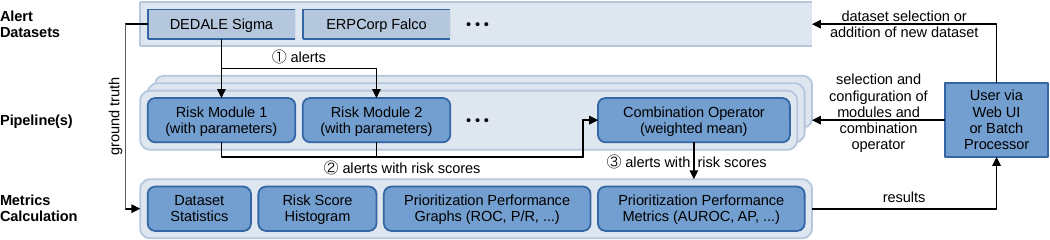}
    \caption{Data flow in \name{}. An alert dataset is processed by one or more pipelines, each containing one or more risk modules.}
    \label{fig:dataflow}
\end{figure*}

\section{Risk Hypotheses for Alert Prioritization}
\label{sec:hypotheses}

We searched literature on RBA for concrete risk incident rules with the goal of identifying their underlying assumptions (i.e., risk hypotheses).
More precisely, we surveyed the Splunk Enterprise Security documentation~\cite{splunk2025how}, Splunk's Guide to Risk-Based Alerting~\cite{splunk2024rba}, the 21 RBA talks referenced therein, and Splunk's RBA repository on GitHub~\cite{splunk2025githubrba}.
We found a total of 12 different risk incident rules, with five of them occurring in multiple sources, as shown in Table~\ref{tab:rirs}.
Additional web searches and large language model queries for risk incident rules revealed no further candidates.

Reviewing the rules, we recognized three abstract schemes:
(1)~the risk scores of alerts are \emph{accumulated} over certain entities and time periods (Rules 1--3);
(2)~the number of \emph{different alert types} is counted for certain entities and time periods (Rules 4--9); and
(3)~events are checked for \emph{first appearance} on certain entities in a specified time period (Rule 10).
Since our goal is to identify and evaluate the fundamental assumptions underlying these rules on how to distinguish true from false alerts, we formulate an explicit risk hypothesis for each of the three schemes:

\textbf{Accumulation:} \emph{Spatio-temporal accumulation of alerts implies higher risk.}
The underlying idea of this hypothesis is that attacks cause an anomalously large number of alerts during a limited time span and related to a limited number of entities.
For example, a port scan against a web server may cause a large number of alerts within a few minutes on that machine, whereas regular activity may cause substantially fewer alerts.
We implement this hypothesis by assigning each alert a risk score that rises with the number of alerts affecting the same entity (e.g., host or user) within a specified time window (cf. Section~\ref{sec:cats}).

\textbf{Variety:} \emph{Spatio-temporal variety in alert types implies higher risk.}
This hypothesis is similar to Accumulation, but instead of counting all close alerts, only the number of \textit{different} triggered detection rules is counted (again with respect to an entity and time span).
The motivation behind this adaptation is that benign misconfigurations or changes in benign activity patterns may cause large numbers of alerts, but usually restricted to just one or very few detection rule(s).
A multi-step attack, on the other hand, is more likely to trigger multiple rules in a limited time frame.
We implement this hypothesis by assigning each alert a risk score that rises with the number of \emph{different} alerts per entity and time window.

\textbf{Rarity:} \emph{Rarely occurring alert types imply higher risk.}
This hypothesis states that attacks deviate from benign activity and are therefore likely to cause alerts that are rare with respect to the triggered detection rule or entity.
Note that this is a generalization of checking for first appearance, with the latter yielding the highest possible risk score.
For example, while a benign application might check for updates and thereby trigger a rule detecting unencrypted HTTP traffic on a daily basis, newly installed malware might trigger a similar rule detecting suspicious FTP traffic for the first time ever.
We implement this hypothesis by assigning each alert a risk score that decreases with the number of occurrences of the same alert type within a specified time window.

In addition to the three above hypotheses, we formulate two further ones not derived from our literature review, but based on straightforward characteristics: the alert severity level and the periodic occurrence of alerts.

\textbf{Rule Level:} \emph{A higher severity level of a detection rule implies higher risk.}
Threat detection rules usually have a level assigned by their author that represents the severity of the detected attack and can thus serve as a risk indicator.
For example, the Sigma format defines five levels from \textit{informational} to \textit{critical}~\cite{sigmarules} that rule authors can choose from.
We include this hypothesis in our evaluation because it constitutes a straightforward approach that is probably used in practice by many SOCs to prioritize alerts and can thus be seen as an implicit baseline.
We implement it by mapping the levels to risk scores between zero and one, e.g., 0.0, 0.25, 0.5, 0.75, and 1.0 for the five Sigma levels.

\textbf{Aperiodicity:} \emph{Aperiodically occurring alerts imply higher risk.}
The assumption here is that automated benign tasks are usually periodic and thus false alerts caused by them are also likely to occur periodically.
For example, administrative cron jobs causing false alerts might be scheduled to run at the same time every day or night, whereas attacks occur irregularly.
While this concept does not seem to be covered in RBA literature (possibly because Splunk's RBA framework does not support periodicity calculation), it is established in the area of anomaly detection~\cite{chandola2009anomaly} and we deem it a natural candidate for evaluation.
We implement this hypothesis by assigning each alert a risk score that decreases with the periodicity of the historic alert type pattern.

Returning to the risk incident rules in Table~\ref{tab:rirs}, note that Rules 11--12 (which do not match any of our five risk hypotheses) are out of scope for our work since the definition of key performance indicators and mean time to resolution are highly specific to an organization and cannot be evaluated generically.
Likewise, note that our alert datasets (cf. Section~\ref{sec:datasets}) do not consistently provide MITRE ATT\&CK~\cite{mitreattack} labels, so we rely on commonly available alert features (i.e., timestamp, source/destination, detection rule ID and severity).

While all five risk hypotheses (Accumulation, Variety, Rarity, Rule Level, and Aperiodicity) may appear sensible, our evaluation will show which of them can actually prioritize alerts well when applied to realistic datasets.
To make this evaluation possible, we now introduce our novel tool \name{}.

\input{Modules}

\section{CATS: Alert Triage Evaluation System}
\label{sec:cats}

\begin{figure}[b]
    \centering
    \includegraphics[width=\columnwidth]{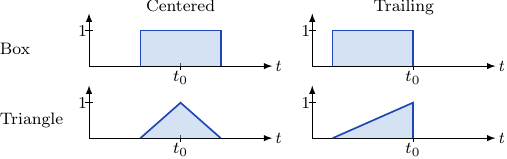}
    \caption{Window alignments and shapes of time-dependent risk modules, where $t_{0}$ is the timestamp of the current alert.}
    \label{fig:slidingwindows}
\end{figure}

To enable a comprehensive evaluation of risk hypotheses against different alert datasets using multiple metrics, we developed \name{}, our \textbf{C}ybersecurity \textbf{A}lert \textbf{T}riage Exploration and Evaluation \textbf{S}ystem.
\name{} allows for either visual exploration (via a web UI, see Appendix~\ref{sec:catsscreenshot} for a screenshot) or automated evaluation (via a batch processor).
\name{} is available on GitHub~\cite{githubcats} and its web UI can be tested on \href{https://962012d09b.github.io/cats_webapp/}{https://962012d09b.github.io/cats\_webapp/}.

Figure~\ref{fig:dataflow} depicts \name{}' data flow, which applies to the web UI as well as the batch processor:
\ding{172}~A selected \textbf{alert dataset} is fed into one or more \textbf{pipelines}, each consisting of one or more parametrized risk modules and a combination operator.
Each \textbf{risk module} implements one risk hypothesis as introduced in the previous section, featuring parameters as listed in Table~\ref{tab:hypotheses}.
These parameters allow to select the entity by which alerts are grouped (e.g., host or user) and the length, alignment, and shape of the applied window function, as visualized in Figure~\ref{fig:slidingwindows}.
We chose the triangle window in addition to the straightforward box window because it smoothly fades out temporally distant alerts but is still finite in length unlike an exponential window, thus avoiding computational issues.
Note that a pipeline may contain multiple instances of a risk module, e.g., with different window lengths.
\ding{173}~Each module instance calculates a \textbf{risk score} $\in [0,1]$ for each alert, where zero denotes a low risk (likely false alert) and one a high risk (likely true alert).
All modules feature a normalization such that their expectation value is approximately 0.5.

The \textbf{combination operator} of a pipeline defines how the overall risk score of each alert is calculated.
It currently supports a weighted arithmetic or geometric mean of the individual module instances' risk scores.
While the arithmetic mean is a natural choice that incorporates all risk scores equally, the geometric mean should be chosen when risk scores are interpreted as probabilities.
However, it should be used with caution since small risk scores near zero strongly decrease the overall risk score, which might not be desired.

\ding{174}~The combined risk scores are then fed into the \textbf{metrics calculation}, producing data for histograms, graphs, and aggregate metrics (cf. Appendix~\ref{sec:catsscreenshot}).
These data are further discussed in Section~\ref{sec:evaluation} and can be stored in JSON format for further processing.
When the batch processor is used, it can be configured to optimize module weights and/or parameters such that a target metric (usually AUROC) is maximized, thus achieving the best possible alert prioritization for the provided dataset, modules, and parameters.

\section{Alert Datasets: Existing and Newly Created}
\label{sec:datasets}

Evaluating the introduced risk hypotheses with CATS requires alert datasets that contain a large number of both true alerts (i.e., induced by attacks) and false alerts (i.e., induced by benign activity) to obtain expressive results.
This rules out attack-only datasets from honeypots~\cite{husak2020dataset} or capture-the-flag events~\cite{meyers2022examining}.
Moreover, datasets should cover diverse environments, attacks, and security systems~\cite{sommer2021outside, lamberts2023sok}.

However, according to a recent survey~\cite{boenninghausen2024comidds} and our own research, there exists only one alert dataset satisfying our requirements (AIT-ADS, see below).
To improve the validity of our evaluation and foster future research, we thus created new, labeled alert datasets from three additional, already-existing sources: (1)~a testbed, (2)~a log dataset, and (3)~an adversary emulation plan.
In addition, we created a proprietary enterprise alert dataset that we are not allowed to share but utilize to strengthen our evaluation.
We make all other datasets publicly available, including code to reproduce them~\cite{githubcats}.

Each dataset is stored in a straightforward JSON Lines~\cite{ward2025jsonlines} format where each line represents one alert, containing the original log record, the triggered rule, and a set of features parsed from the log record or rule.
In addition, each alert is labeled as either true (attack) or false (benign).
See Appendix~\ref{sec:catsalert} for an exemplary Sigma alert in the CATS format.

We briefly describe the five sources for our eight alert datasets in the following.
In addition, Table~\ref{tab:datasets} shows key characteristics of the datasets, where ``Alert Types'' denotes the number of distinct triggered detection rules and ``Duration'' denotes the time delta between the first and last alert (which may be shorter than the actual recording time).
Note that we describe the dataset sources in ascending order of our contribution (preexisting to entirely created by us) but sort the resulting datasets by duration throughout the rest of this paper to roughly represent their expressiveness in descending order (cf. Table~\ref{tab:datasets}).
Furthermore, note that we separate datasets by the security system that generated the alerts, e.g., we refer to AIT-ADS \emph{Wazuh} and AIT-ADS \emph{Suricata} as two datasets even though they stem from the same scenario.
We do this because host-based and network-based alerts require different parametrization (cf. Section~\ref{sec:evaluation}).

\textbf{AIT-ADS} (Austrian Institute of Technology Alert Data Set) is based on the AIT Log Data Set V2.0~\cite{landauer2023maintainable, landauer2022aitlds2} and was published in 2024~\cite{landauer2024ads, comidds2024ads}.
The underlying testbed models a small company network with 9--27 Ubuntu Linux clients, several DMZ servers, and a simulated Internet zone.
Benign user activity is implemented via state machines.
The simulated multi-step attack scenario incorporates steps such as webshell upload, password cracking, and DNS exfiltration.

AIT-ADS contains Wazuh, Suricata, and AMiner~\cite{landauer2023aminer} alerts.
We omit the latter since they miss required fields such as hostname and rule level.
The dataset comprises eight scenarios with the same environment and attack cases but variations in parametrization.
We chose the scenario with the fewest alerts (``russellmitchell'') since CATS can process it in a few seconds, allowing for responsive visual exploration.

\input{Datasets}

\textbf{SOCBED} is a testbed focusing on reproducible and adaptable log data generation, published in 2021~\cite{uetz2021socbed, comidds2024socbed}.
It models a small company network with three zones (internal, DMZ, Internet), Windows clients, and common services such as web, mail, and directory.
An agent running on each client emulates benign user activity, namely, web surfing, emailing, and file operations.
The testbed provides a multi-step cyber espionage attack, including web server exploitation, information gathering, and backdoor installation.

We executed a two-hour SOCBED simulation with an eight-step attack starting after one hour and lasting for approximately 30 minutes.
We then collected the resulting Windows event logs and BSD syslogs.
We ran Sigma rules against the former, extracted Suricata alerts from the latter, and finally labeled all alerts manually.

\textbf{DEDALE} is a ``Dataset for Evaluating Detection of APT among Logs and Events'', published in 2025~\cite{lanvin2025dedale, lanvin2024dedalewebsite, comidds2024dedale}.
Its underlying testbed RESCOUSSE is a substantial extension of SOCBED and models a corporate network and external systems with a total of 55 virtual machines (30~of them being Windows clients), divided into four network zones.
Benign users with different roles (e.g., manager or developer) are simulated, including activities such as web browsing, emailing, file operations, etc.
An APT attack is carried out over eight days, comprising eight ATT\&CK tactics.
DEDALE contains labeled Windows event logs, Linux syslogs, and network logs.

We created an alert dataset from these raw log data by running Sigma rules against the Windows event logs, carefully reviewing, and partially correcting the associated labels to achieve a trustworthy ground truth.
Note that DEDALE also contains Suricata alerts, but since only two alerts out of 130k are attack-induced, we omitted them to avoid highly unstable metrics and thus potentially misleading results.

\textbf{APT29S2} denotes an APT29~\cite{mitre2025apt29} emulation plan (``Scenario 2'') provided by MITRE~\cite{github2025apt29}, which we picked due to its realistic simulation of activity from an existing state-sponsored threat group.
We built an environment as required for this scenario, consisting of a Windows domain controller and two Windows workstations.
We used the GHOSTS framework~\cite{github2024ghosts} to simulate benign activity on the workstations (web browsing, file modifications, and PowerShell commands) and administrative activity on the domain controller (file modifications and PowerShell commands).

We executed the attack steps using the adversary emulation tool Caldera~\cite{mitre2024caldera}, including gathering domain information, escalating privileges, and forging a Kerberos golden ticket~\cite{qatinah2024kerberos}.
After the emulation, we ran Sigma rules against the resulting Windows event logs and Suricata rules against the captured network traffic.
As with the other datasets, we carefully annotated all resulting alerts as true or false.

\textbf{\datasetname} denotes a proprietary dataset that we recorded in the production environment of a corporation as part of this work.
The company operates Linux-based application servers to provide its enterprise resource planning (ERP) software to customers.
Benign activity comprises customers using this software as well as administrative tasks (e.g., software orchestration).
We simulated a realistic multi-step attack where an adversary gains initial access to an application server, escalates privileges using a Docker socket, installs a persistent reverse shell backdoor using a systemd unit, collects confidential files, exfiltrates them via \texttt{scp}, and finally runs a file-encrypting ransomware.

Host-based data (i.e., syscall logs) were collected by agents running on 44 ERP application servers. Alerts were generated by a company-specific Falco~\cite{falco2026falco} rule set.

Using the eight obtained alert datasets (cf. Table~\ref{tab:datasets}), we can now evaluate the five introduced risk hypotheses (cf. Section~\ref{sec:hypotheses}) using CATS (cf. Section~\ref{sec:cats}).

\input{Heatmap}

\section{Evaluation of Risk Hypotheses}
\label{sec:evaluation}

According to practical experience, risk-based alerting can massively help in finding more true alerts while at the same time having to review fewer false alerts (cf. Section~\ref{sec:hypotheses}).
However, until now, there has neither been a sound quantitative evaluation of this claim nor an actionable set of experiments to figure out good (or bad) risk hypotheses along with their parameters and potential combinations.
We break this issue down into three research questions:

\textbf{RQ1:} \emph{Which risk hypotheses prioritize alerts well and robustly?}
To investigate, we measure the prioritization performance of the five hypotheses across all datasets and discuss the influence of parameters on the results (Section~\ref{sec:rq1}).

\textbf{RQ2:} \emph{Can a combination of risk hypotheses prioritize alerts even better?}
To address this question, we determine the optimal combination of hypotheses for each dataset and then perform a leave-one-out cross validation (Section~\ref{sec:rq2}).

\textbf{RQ3:} \emph{What other metrics are relevant in practice?}
Finally, we discuss multiple metrics that each answer specific questions and thus offer different perspectives on prioritization performance for practical application (Section~\ref{sec:metrics}).

\subsection{Three Modules Prioritize Alerts Well}
\label{sec:rq1}

First and foremost, we investigate if the individual risk modules (each implementing one risk hypothesis) can achieve an alert prioritization performance above chance level and how these results differ across datasets and parameters.

\textbf{Methodology.}
We used the \name{} batch processor to apply each risk module with all sensible parameter combinations to each of our eight alert datasets:
For Rule Level, which has no parameters, only one instance was applied.
For Accumulation, Variety, Rarity, and Aperiodicity, which have similar parameters, we applied all combinations of the following parameters.
\emph{Group By} was set to \emph{Hostname} for host-based datasets, \emph{Source IP} and \emph{Destination IP} for network-based datasets, and \emph{Alert Type} for both.
We omitted \emph{Username} as a parameter since this field is missing in most alerts of our datasets.

Concerning window length, we picked \emph{one minute}~(1m), \emph{one hour}~(1h), and \emph{one day}~(1d).
We did not evaluate longer windows (e.g., \emph{one week}) since most datasets are shorter.
Both \emph{centered} and \emph{trailing} windows were tested, as well as \emph{triangle-} and \emph{box-}shaped windows.
Two exceptions apply: (1) Variety does not support \emph{Group By Alert Type} since it always counts unique alert types, and (2) Aperiodicity does not support a triangle window shape since it is undefined for the underlying Fourier transform (cf. Table~\ref{tab:hypotheses}).

We deliberately chose AUROC as our primary evaluation metric because it rates prioritization performance uniformly, i.e., independent of alert position.
This makes sense for our setting since SOCs differ in alert volume and analyst resources and consequently in their ratio of triaged to total alerts, rendering top-heavy metrics such as average precision too specific (see Section~\ref{sec:metrics} for AP results).

\textbf{Results.}
Table~\ref{tab:heatmap}~(a) visualizes the evaluation results for each risk module, window length, and dataset.
Each colored cell shows the AUROC value of the respective module applied to the respective dataset.
AUROC arithmetic means and standard deviations across the eight datasets are shown on the right, together with the number of datasets exceeding chance level.
As a baseline, no prioritization equals an AUROC of 0.5 (chance level).

The results show that Rule Level, Accumulation, and Variety prioritize alerts well above chance level for almost all datasets, whereas Rarity and Aperiodicity achieve mediocre results with rather high standard deviations (see Appendix~\ref{sec:notableresultsappendix} for a discussion of notable individual results).
The best modules with respect to mean are Variety 1h and 1d (performing above chance level for all datasets), followed by Accumulation 1m.
Rule Level also performs well overall, but is close to chance for three out of eight datasets.

We can see that the modules have different optimal window lengths:
Accumulation works best for a short 1m window, Variety is strongest at 1h, and Rarity and Aperiodicity increase in performance with longer windows.
An interpretation could be that Accumulation best detects short bursts of attack-induced alerts, Variety discovers multiple different attack steps over medium to long periods, and Rarity and Aperiodicity best capture deviations from longer-term benign activity.

The above results were generated using the parameters shown as default options in Table~\ref{tab:hypotheses} (with \emph{Group By} in Accumulation and Variety set to \emph{Hostname} for host-based datasets and \emph{Source IP} for network-based datasets).
Next, we discuss alternative parametrizations.
Note that all mentioned results can be easily reproduced with \name{}~\cite{githubcats}.

Using \emph{Destination IP} instead of \emph{Source IP} had an overall slightly negative effect with respect to the affected modules (Accumulation and Variety) on network-based datasets ($\mu=0.61$ vs. $0.65$, $\sigma=0.24$ vs. $0.28$).
Interestingly, we found that AIT-ADS Suricata works better with \emph{Source IP}, APT29S2 Suricata with \emph{Destination IP}, and for SOCBED Suricata, there is almost no difference.
We thus recommend to try both parameters when using own datasets.

Similarly, switching \emph{Hostname}/\emph{Source IP} with \emph{Alert Type} (Accumulation) and vice versa (Rarity, Aperiodicity) leads to an overall worse prioritization ($\mu=0.41$ vs. $0.59$, $\sigma=0.17$ vs. $0.18$ w.r.t. the affected modules).
We expected these results since the switched parametrization does not reflect the original idea of the hypotheses anymore.
In particular, the switched Accumulation module behaves similarly to an inverted Rarity module with default parameters and vice versa, making this configuration dispensable.

Furthermore, we compared window alignments (centered or trailing) and shapes (triangle or box), as visualized in Figure~\ref{fig:slidingwindows}.
With a trailing instead of centered window, Rarity and Aperiodicity preserve their performance while Accumulation and Variety become worse ($\mu=0.70$ vs. $0.79$, $\sigma=0.05$ vs. $0.06$).
This seems natural since the latter modules benefit from incorporating all surrounding alerts, not just preceding ones.
Comparing triangle to box windows, the difference is marginal ($\mu=0.79$ for both, $\sigma=0.07$ vs. $0.06$).
However, the box window bears the risk of abrupt risk score changes when an alert cluster slides out of the window, thus we recommend using the triangle window.

\textbf{Summary.}
Our evaluation of risk modules across all datasets indicates that Variety, Accumulation, and (to a lesser extent) Rule Level prioritize alerts well and robustly.
While Accumulation works best for the short one-minute window, Variety benefits from medium (one hour) to long (one day) windows.
The best overall parametrization is shown as default in Table~\ref{tab:hypotheses}, but differences are modest.

\subsection{Combining Modules Works Even Better}
\label{sec:rq2}

\begin{figure}[t]
    \centering
    \includegraphics[width=\columnwidth]{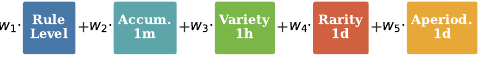}
    \caption{We define a pipeline as a weighted combination (sum or product) of risk modules and optimize the weights to evaluate its performance advantage over single modules.}
    \label{fig:pipeline}
\end{figure}

Since each risk module implements a different prioritization hypothesis, the question arises whether they could complement each other and thus achieve a prioritization performance beyond that of the best single module.

\textbf{Methodology.}
To address this question, we combined the risk scores of all modules into a weighted mean.
More precisely, we created a CATS pipeline containing each risk module with its best-performing window length, as depicted in Figure~\ref{fig:pipeline}.
Using the batch processor, we numerically optimized the module weights with respect to the AUROC of the weighted mean.
Note that optimizing longer pipelines, e.g., with multiple differently parametrized instances per risk module, was computationally infeasible due to a combinatorial explosion of the weights.
For the same reason, we restricted module weights to steps of 0.1 and their sum to 1.0, resulting in a total of 1001 evaluated weight combinations per dataset.
Consequently, we obtained the best-performing module weights for each dataset.

However, these results are not valid for predicting performance on unseen data since they are optimized on the data itself.
We thus performed a leave-one-out cross validation: For each dataset, we tested the performance of a pipeline with module weights set to the arithmetic mean of the optimized weights of \emph{all other} datasets, thereby avoiding data snooping~\cite{arp2022dos}.
Note that we also utilized leave-one-out to determine the window lengths of the pipeline modules (cf. Figure~\ref{fig:pipeline}) for the same reason, but they turned out to be the same for all datasets.
We repeated the whole process with alternative parameters, i.e., window alignment, window shape, and combination operator (cf. Section~\ref{sec:cats}).

\textbf{Results.}
Similar to Section~\ref{sec:rq1}, we first present results for the default module parameters (cf. Table~\ref{tab:hypotheses}) and then discuss parameter variations.
Table~\ref{tab:stacked} shows the results of the numerical weight optimization.
In the middle column, we can see that the optimal module weights differ substantially between datasets.
On average, Variety 1h has the highest weight whereas Rarity 1d and Aperiodicity 1d have the lowest weight, correlating with their lower prioritization performance.
Interestingly, all results include at least two modules, backing our assumption that modules can complement each other.
The resulting performance for this intra-dataset optimization is shown in Table~\ref{tab:heatmap}~(b).
Inherent to the optimization method, each AUROC score is at least as good as that of the best single module for the respective dataset.

The right column of Table~\ref{tab:stacked} shows the leave-one-out module weights for each dataset, i.e., the average of all other datasets' weights with the purpose of predicting prioritization performance on unseen data.
Due to the averaging, the weights are mostly similar across datasets.
As can be seen in Table~\ref{tab:heatmap}~(b), the leave-one-out pipelines achieve a high AUROC mean of 0.92 and a low standard deviation of 0.09, indicating that this module combination seems to work well for diverse environments, alert sources, and attacks.

\input{Stacked}

To give a practical example of the pipeline performance, consider a SOC where analysts are confronted with an overwhelming number of alerts and can only review the highest-prioritized 10\%.
For the AIT-ADS Suricata dataset, which has the lowest (and thus most challenging) base rate (44 true, 9142 false alerts), its leave-one-out pipeline places 43 true alerts in the top 10\% (thus missing only one), whereas a prioritization by rule level would only reveal 28.6 true alerts on average, missing 15.4.
For the DEDALE Sigma dataset, which has the second-lowest base rate (51 true and 1768 false alerts), its leave-one-out pipeline places 50 true alerts in the top 10\% (thus also missing just one), in contrast to Rule Level with only 7.7 true alerts on average.

Next, we examined the impact of parameter changes on the optimized leave-one-out pipelines (see Table~\ref{tab:heatmap}~(c)).
Using a geometric instead of arithmetic mean has a marginal effect.
Switching from triangle to box window causes a 5\% decrease in mean AUROC, and changing the centered to a trailing window results in a 17\% drop and considerably larger standard deviation as compared to the default parameters.
Finally, we evaluated the effects of leaving out (1)~the Rarity and Aperiodicity modules due to their mediocre performance (see Appendix~\ref{sec:altpipeappendix}) and (2)~the SOCBED- and APT29S2-based datasets due to their short duration (see Appendix~\ref{sec:altoptappendix}), finding only minor performance changes.

\textbf{Summary.}
We have shown that a weighted mean of risk modules can achieve high prioritization performance across all datasets (AUROC $\mu=0.92$, $\sigma=0.09$), substantially beyond a straightforward prioritization by rule level.
These results indicate that risk-based alerting, as implemented in this work, might indeed be able to mitigate cybersecurity alert fatigue, aligning with various practitioner reports.

\subsection{Digging Deeper with Multiple Metrics}
\label{sec:metrics}

\begin{figure*}[t]
    \centering
    \includegraphics[width=0.97\textwidth]{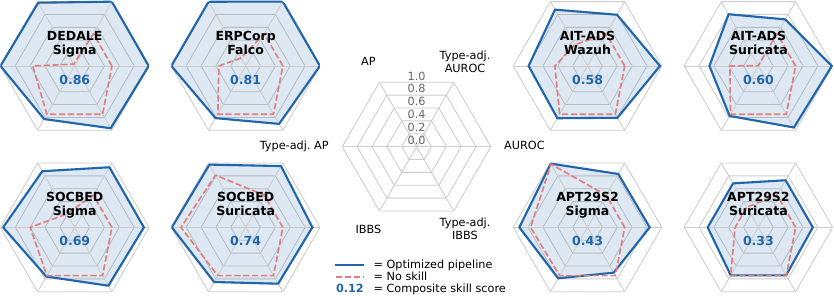}
    \caption{Diverse performance metrics for the optimized leave-one-out pipelines across the eight datasets. The vast majority of the results are substantially above no-skill level, indicating a remarkable alert prioritization performance on unseen data.}
    \label{fig:radarplots}
\end{figure*}

Up to this point, we solely utilized AUROC to rate risk modules and their combinations.
While we deem AUROC to be the best metric for overall prioritization performance in our setting (cf. Section~\ref{sec:rq1}), it does not capture all aspects relevant for practical application~\cite{lamberts2023sok}.

\textbf{Methodology.}
To facilitate a deeper understanding of the risk modules' prioritization characteristics, we selected two additional metrics that shed light on different performance aspects.
The first one is average precision (AP), representing the area under the precision-recall curve, which penalizes highly-ranked false alerts
much stronger than lower-ranked ones.
Considering this metric is advisable if analysts can only review a small fraction of alerts.

Our second additional metric is a variant of the Brier score~\cite{gneiting2007strictly}, which measures calibration:
A perfectly-calibrated module or pipeline would always assign a risk score of one to true alerts and zero to false alerts.
While AUROC and AP solely consider the order of risk scores but not their absolute values, the Brier score expresses to which extent risk scores equal attack probabilities, which can be helpful to analysts.
We invert the Brier score to match the other metrics' meaning of ``larger equals better'' and balance it by assigning equal weight to true and false alert calibration.
Balancing is sensible because otherwise a trivial module that always assigns zero would achieve a near-perfect score for datasets with a low base rate, which we find undesirable.
We define the inverse balanced Brier score as \( \mathrm{IBBS} = 1 - \tfrac{1}{2} \left( \mathrm{MSE}_T + \mathrm{MSE}_F \right) \), where \( \mathrm{MSE}_T \) and \( \mathrm{MSE}_F \) are the mean squared errors of the risk scores of the true and false alerts, respectively.

We noticed that for some datasets, a small fraction of alert types constitutes the majority of alerts.
Consequently, all previous metrics inadequately express the prioritization performance on less frequent types.
We therefore define a variant of each metric that is adjusted for alert type prevalence, thus giving equal weight to each alert type within a dataset.
Comparing the unadjusted to the adjusted metric can reveal issues in prioritizing certain alert types.

Note that we deliberately omit the commonly used metrics F1 score and Matthew's correlation coefficient (MCC) since they only represent the performance at one threshold (i.e., share of reviewed alerts) whereas each SOC has a different and likely varying threshold.
For detailed inspection, \name{} allows to plot F1 and MCC across all thresholds.

In addition to the six metrics described above, we introduce a new score that summarizes all of them into one number and is meant to give a quick impression of the overall performance of a module or pipeline.
This score measures the mean exceedance of the \emph{no-skill level} across the six metrics, with the no-skill level being the performance of the \emph{best} classifier with no discriminative ability.
Thus, the no-skill level for AUROC is 0.5, for AP it is the dataset base rate, and for IBBS it is 0.75 (achieved by a classifier always returning~0.5).
Note that IBBS is the only one of these metrics where the no-skill level is not achieved by random classifiers (e.g., a coin-flip classifier yields a score of 0.5, making 0.75 a rather demanding no-skill level for risk modules or pipelines).

We define the \emph{skill score} of a metric as the score exceeding no-skill level, scaled such that no skill\,=\,0 and perfect\,=\,1.
Based on that, we define the \emph{composite skill score} (CSS) as the mean of the skill scores of the six metrics.

\textbf{Results.}
Figure~\ref{fig:radarplots} shows the six described metrics for the optimized leave-one-out pipelines across all datasets, along with the respective no-skill levels and composite skill scores.
Almost all metrics exceed no-skill level, the only lower value is the type-adjusted IBBS for APT29S2 Sigma, indicating a mediocre risk score calibration on this dataset.
Overall, AUROC and AP scores (both unadjusted and adjusted) are high, even for challenging datasets with a low base rate.
This is also reflected in the CSS values, which are all substantially larger than zero (i.e., no-skill level).

\textbf{Summary.}
Additional metrics can provide a more complete picture of prioritization performance.
For the optimized pipelines, the six utilized metrics show mostly high values across datasets with respect to no-skill level and no critical failures, indicating fitness for practical application.

\section{Discussion and Limitations}
\label{sec:discussion}

We now look at our results from a high-level perspective and discuss potential limitations of our approach and datasets as well as considerations for practical application.

\textbf{Discussion of Key Findings.}
First and foremost, we consider it remarkable how well some of the risk modules (and especially their combination) prioritize alerts, given that they use only straightforward statistics over a small number of common alert features.
The practical example given in Section~\ref{sec:rq2} indicates that SOC efficiency can be substantially improved by using the optimized pipeline.
More specifically, our results show that Rule Level, Accumulation, and Variety can distinguish true from false alerts significantly above chance level (one-sided Wilcoxon signed-rank test against $\mathrm{AUROC}=0.5$: $p=0.016$, 0.012, and 0.0039 for the default configuration, respectively).
This is not the case for the Rarity ($p=0.63$) and Aperiodicity ($p=0.42$) modules.
Interestingly, the two best performers -- Accumulation and Variety -- are also the ones most strongly represented in practical RBA literature (cf. Table~\ref{tab:rirs}), confirming their broad applicability.

\textbf{Caveats of Our Approach.}
Two implementation decisions should be kept in mind.
First, we present results primarily based on AUROC scores, which we deliberately chose since they cover all possible alert triage rates (cf. Section~\ref{sec:rq1}).
If this rate is known for a concrete SOC, prioritization performance could be rated specifically at this point, e.g., by inspecting the precision and recall graphs in \name{}.

Secondly, our evaluation indicates that a centered sliding window performs better for Accumulation and Variety than a trailing one (cf. Figure~\ref{fig:slidingwindows}).
However, using the former means that alerts cannot be prioritized instantly when they occur but only after a delay of half the window length (e.g., 30 minutes).
While it is common for risk incident rules (cf. Section~\ref{sec:background}) to be executed periodically (e.g., once per hour or once per day) to discover temporal attack patterns~\cite{splunk2024rba}, SOCs requiring real-time prioritization can use the trailing window at manageable performance cost (cf. Table~\ref{tab:heatmap}~(c)).

\textbf{Limitations of the Evaluation Datasets.}
Despite our efforts to utilize diverse datasets, there are some limitations worth mentioning.
First, only \datasetname{} Falco contains real benign activity from a corporate network, the rest stems from testbeds with simulated users.
Yet, the results on this dataset are among the best of all datasets, indicating that the synthetic ones are not per se easier to prioritize.

Secondly, the datasets based on SOCBED and APT29S2 have a rather short duration (cf. Table~\ref{tab:datasets}), rendering the respective results less meaningful for larger window lengths (cf. Appendix~\ref{sec:altoptappendix}).
Ideally, each dataset should span several weeks such that one-week windows could be soundly evaluated.
However, simply running the underlying testbeds longer would likely not be sensible because the simulated user activity is relatively simple and thus quickly becomes repetitive, possibly making alert prioritization too easy.

Thirdly, while environments, benign activity, and attacks differ between datasets, there is an emphasis on advanced persistent threat-like attacks~\cite{attackgroups} as well as open-source threat detection systems (cf. Section~\ref{sec:background}).
More diversity with respect to attacks and detection systems would be desirable but is difficult to implement, especially for datasets that can be made publicly available.

\textbf{Attacks Against RBA.}
Adversaries who are aware that RBA is used to prioritize alerts might aim to exploit it to evade detection.
To this end, they could try to decrease risk scores of alerts they caused -- for example, by performing attack steps with time gaps exceeding the window lengths of the Variety and Accumulation modules.
Alternatively, attacks could be performed periodically to disguise them as benign activity and evade the Rarity and Aperiodicity modules.
However, evasion is not quite as easy because single attack steps commonly trigger multiple detection rules at once, complicating evasion of Variety and Accumulation.
Furthermore, the first appearance of a periodic attack would still cause a high-risk alert and thus likely lead to an investigation.

Another evasion strategy could be obfuscating the actual activity by deliberately causing unrelated, high-risk alerts.
For example, adversaries might launch large-scale attacks against a web server while simultaneously performing lateral movement in an infiltrated network.
However, such deception maneuvers are equally effective when no RBA is employed, thus not being an argument against it.
Overall, while attackers might partially succeed in evading RBA, the results will most likely still be better than without it.

\textbf{Human Factors.}
Our work shows that RBA can prioritize alerts well and thus substantially reduce the number of false alerts that analysts have to triage.
It is thus reasonable to assume that implementing RBA in a SOC ultimately reduces cybersecurity alert fatigue (which is largely caused by false alerts~\cite{alahmadi2022false}).
However, there are additional components to alert fatigue that go beyond the scope of this paper, e.g., inadequate alert explainability~\cite{alahmadi2022false}, non-integrated secondary information sources~\cite{nobles2022stress}, or poorly designed user interfaces~\cite{franklin2017toward}.
Even within the scope of RBA, auxiliary factors might contribute to alert fatigue.
For example, analysts might not trust the prioritization and still check low-priority alerts, thus potentially decreasing RBA's value.
Similarly, while reducing workload, RBA adds a layer of complexity that analysts might need to get acquainted with initially.
Future work should thus conduct user studies with SOC analysts to directly measure the effect of RBA on cybersecurity alert fatigue.

\textbf{Considerations for Practical Application.}
As described in Section~\ref{sec:problem}, our RBA implementation differs from Splunk's in that we prioritize the original alerts instead of generating non-prioritized notables, thus allowing us to evaluate all possible alerting thresholds at once.
Still, our results apply to Splunk's approach because notables can be seen as a special case of our approach.
Consider a risk incident rule that generates a notable when three different alert types occur on a host within one day.
By contrast, our Variety module continuously increases alert risk scores with the number of different alert types and their temporal distance (when using the triangle window), thus also covering this special case.

Alternatively, our prioritization-based approach could also be used in practice.
To this end, the risk score can simply be added as a column to the table showing alerts for manual review.
Analysts could then either sort alerts by timestamp (as without RBA) or by risk score, allowing them to review alerts in descending order of risk.
This approach allows for a smooth transition to RBA because analysts can easily switch between the temporal and risk-based views.

Finally, to make RBA more transparent, we would suggest to store and display the risk scores of all pipeline modules individually so that analysts can see which risk hypotheses primarily contributed to the overall risk score.

\textbf{Use of RBA as a Baseline.}
A major benefit of RBA is its low computational complexity.
Specifically, we measured an average \emph{single-thread} throughput of 413 alerts per second (35.7 million per day) across all datasets on a desktop PC (AMD Ryzen~9 7950X3D CPU, 64\,GB RAM) using our five-module pipeline in CATS (cf. Section~\ref{sec:rq2}).

Since parallelization would be straightforward to implement, the approach is almost certainly fast enough to run on one commodity server even in large enterprise networks.
Consequently, more computationally expensive approaches (especially those involving machine learning, including large language models) should be evaluated against RBA as a baseline and clearly demonstrate that their additional cost is justified by improved prioritization performance.

\section{Related Work}
\label{sec:relatedwork}

There exists a large body of work on reducing cybersecurity alert fatigue.
A recent survey divides this field into three subfields: \emph{Automation} (comprising alert prioritization, false alert reduction, capacity \& workload management, and data triage), \emph{augmentation} (comprising visualization, explanation, and sonification), and \emph{collaboration} (comprising active learning and interactive conversational interfaces)~\cite{shahroz2025alertfatigue}.
In keeping with the focus of our paper, we restrict our discussion to works targeting alert prioritization~\cite{jalalvand2024alert} and false alert reduction~\cite{huballi2014survey}.
Note that within each of the following paragraphs, we discuss the referenced works in chronological order (oldest to newest), since ordering them by relevance with respect to our work is not straightforward.

The methods proposed in the above two subfields can be further divided by their required input data:
While the approach described in this work solely builds on alerts of commonly deployed security systems, various methods require additional data, namely, system-level telemetry~\cite{hassan2019nodoze, hassan2020tactical, liu2022rapid, sharif2024drsec}, network packet captures~\cite{jajodia2026before}, asset information (e.g., business criticality)~\cite{anuar2011risk, alsubhi2012fuzmet, shah2019twostep}, continuous analyst feedback~\cite{doak2013active, renners2019feedback, liu2022context, wang2024combating, turcotte2025aact}, or labeled training data~\cite{sopan2018building, ndichu2021machine, gelman2023teq}.

Since approaches requiring additional data are not directly comparable to ours, we limit the following discussion to methods that require only alerts as input.
We begin with three works that focus on alert correlation and address prioritization only peripherally.
Chyssler et al.~\cite{chyssler2004alarm} combine filtering, aggregation, and correlation to reduce alerts.
All steps involve binary decisions and thus cannot be used for alert prioritization.
\textsc{DeepCase}~\cite{vanede2022deepcase} correlates alert sequences from the same device to support analysts.
The approach does not feature autonomous prioritization but derives priorities from analyst-labeled alerts.
MATE~\cite{lin2022mate} correlates alerts into cases based on ATT\&CK tactics.
Case prioritization is based on alert scores and the number of covered tactics, but only tested for two exemplary cases.

The following works primarily focus on alert prioritization or false alert reduction and are thus most closely related to our work.
Spathoulas et al.~\cite{spathoulas2010reducing} propose an alert filter that is similar to our Accumulation module.
However, it depends on source and destination IPs and is thus not applicable to host-based alerts.
Moreover, an attack-free calibration window is assumed, which might be difficult to ensure.
Zomlot et al.~\cite{zomlot2011dempster} build hypotheses such as ``machine is compromised'' from Snort~\cite{snort} alerts and perform prioritization via Dempster-Shafer belief propagation.
While the approach is promising and might be implemented as a \name{} module, it requires mapping Snort rules to hypotheses, which is not trivial to perform and specific to this alert source.

OutMet~\cite{shittu2014outmet} groups alerts into correlation graphs based on feature similarity and prioritizes them by their difference to neighboring graphs.
Similar to Spathoulas et al., the approach is designed for network-based alerts and is not trivially applicable to host-based systems.
Finally, the work of Kapera et al.~\cite{kapera2025dynamic} is the only one that explicitly addresses risk-based alerting.
It proposes a concept to dynamically adapt risk thresholds for alerting depending on temporal risk changes and differences between user groups to reduce false alerts.
Our work avoids this dependence on alerting thresholds by formulating RBA as a prioritization problem where analysts review alerts in descending risk order until their (temporal) resources are depleted (cf. Section~\ref{sec:problem}).

Note that the vast majority of cited works base their evaluations on either proprietary or meanwhile obsolete datasets (DARPA 1999~\cite{lincoln1999darpa, lippmann2000darpa} and LLDOS~\cite{lincoln2000darpa, haines2001extending}).
Exceptions only exist within the works requiring additional data (e.g., Turcotte et al.~\cite{turcotte2025aact} use AIT-ADS~\cite{landauer2024ads}).
In contrast, our evaluation is based on eight recent datasets that are in part created/extended and published by us (cf. Section~\ref{sec:datasets}).
Furthermore, to the best of our knowledge, we are first to introduce a modular experimentation suite that allows direct comparisons of risk hypotheses and parameters as well as other methods, which can be added as new modules.
We thus encourage researchers to utilize our datasets, software, and results to develop even better alert prioritization methods.

\section{Conclusion}
\label{sec:conclusion}

This paper presents the first systematic evaluation of risk-based alerting, a promising methodology for improving cyberattack detection and reducing alert fatigue in security operations centers.
To investigate the utility of RBA, we formulated the process of alert triage as a prioritization problem and tested five risk hypotheses for their capability to rank true alerts higher than false alerts.
Our evaluation across eight alert datasets shows that three hypotheses succeed in doing so, namely, (1) spatio-temporal variety in alert types, (2) spatio-temporal accumulation of alerts, and (3) high severity levels of alerts.
By contrast, ranking based on the rarity or periodicity of alert types did not significantly exceed chance level.
Most notably, a weighted combination of all hypotheses achieved an even higher mean prioritization performance, with lower standard deviation, than the best single hypothesis.

Our results indicate that RBA can substantially reduce the number of false alerts that analysts need to review, thus increasing SOC efficiency.
In particular, the combination of all tested hypotheses appears to generalize well to diverse datasets and is relatively robust with respect to parametrization.
We thus recommend security operations centers struggling with too many false alerts to try RBA and apply the insights presented in this work.
To lower the entry barrier, practitioners and researchers can employ our experimentation suite CATS to explore and tune RBA performance, optionally using their own alert data.
In addition, we recommend RBA as a baseline for more complex, resource-intensive alert prioritization methods (e.g., large language models).
Moreover, we see potential in implementing further risk modules in CATS.
This would enable direct comparisons of different approaches and might even reveal novel synergies.

In conclusion, we find that risk-based alerting \emph{can} indeed mitigate cybersecurity alert fatigue, at least under the assumption that analysts benefit from having to review substantially fewer false alerts (cf. Section~\ref{sec:discussion}).
Our work shows a concrete path to implementing RBA in security operations centers as well as directions for further research and thus contributes to an improved detection of cyberattacks.

\section*{Open Science}

We provide an anonymized version of the CATS code, alert datasets, and documentation for verifying, reproducing, and extending our results at \href{https://github.com/962012d09b/cats}{https://github.com/962012d09b/cats}.
This repository will be deanonymized and made publicly available in case of paper acceptance.
It is structured as follows:

\begin{tightemize}
    \item \texttt{README.md} provides a brief introduction to the CATS web UI and instructions for hosting an own backend. %
    \item \texttt{backend/} contains the CATS Python backend used by both the web UI and the batch processor (cf. Section~\ref{sec:cats}), including all of the introduced risk modules (cf. Section~\ref{sec:hypotheses}).
    \item \texttt{datasets/} contains the alert datasets used in this paper (cf. Table~\ref{tab:datasets}) in zipped JSON Lines format, except for ERPCorp Falco, which we are not permitted to publish due to security and privacy concerns of the source corporation.
    \item \texttt{docs/} contains a definition of the alert dataset format and data model for the purpose of creating additional datasets.
    \item \texttt{frontend/} contains the CATS web UI code.
    \item \texttt{tools/} contains (1)~the CATS batch processor (cf. Section~\ref{sec:cats}) and (2)~scripts for reproducing our alert datasets from the original datasets (cf. Section~\ref{sec:datasets}).
\end{tightemize}

In addition, we provide a \emph{hosted} instance of CATS (which can be tested without the need for a local installation) at \href{https://962012d09b.github.io/cats_webapp/}{https://962012d09b.github.io/cats\_webapp/}.

Using this URL, the alert prioritization metrics presented in this paper can be reproduced within minutes.
Please note that this CATS instance might not work correctly when accessed through a web proxy since its backend uses a non-standard port that may get blocked by some proxy servers.
Also note that it is a read-only instance that does not permit making persistent changes such as saving pipeline configurations on the server.
Still, downloading results and importing/exporting pipeline configurations in JSON format is fully supported.

\section*{Ethical Considerations}

The evaluation in this work is based on multiple datasets, of which one (ERPCorp Falco) warrants discussion because it was recorded in a productive enterprise network with real users, whereas the others are based on testbed simulations that, to our knowledge, contain no personal information.
We created ERPCorp Falco in the context of evaluating host-based threat detection for the corporation.
The approach, procedure, and use of the dataset were discussed with and approved by its head of information security.
Access to the dataset is restricted to the authors of this paper; none of its contents were ever uploaded to external services.
While the dataset contains internal IP addresses as well as names of processes and files, we comprehensively examined all values and could not find any personally identifiable information since it solely captures system-level activity as well as some administrative commands executed on servers.

\bibliographystyle{plainurl}
\bibliography{literature.bib}

\appendix

\begin{figure*}[p]
    \centering
    \includegraphics[width=\textwidth]{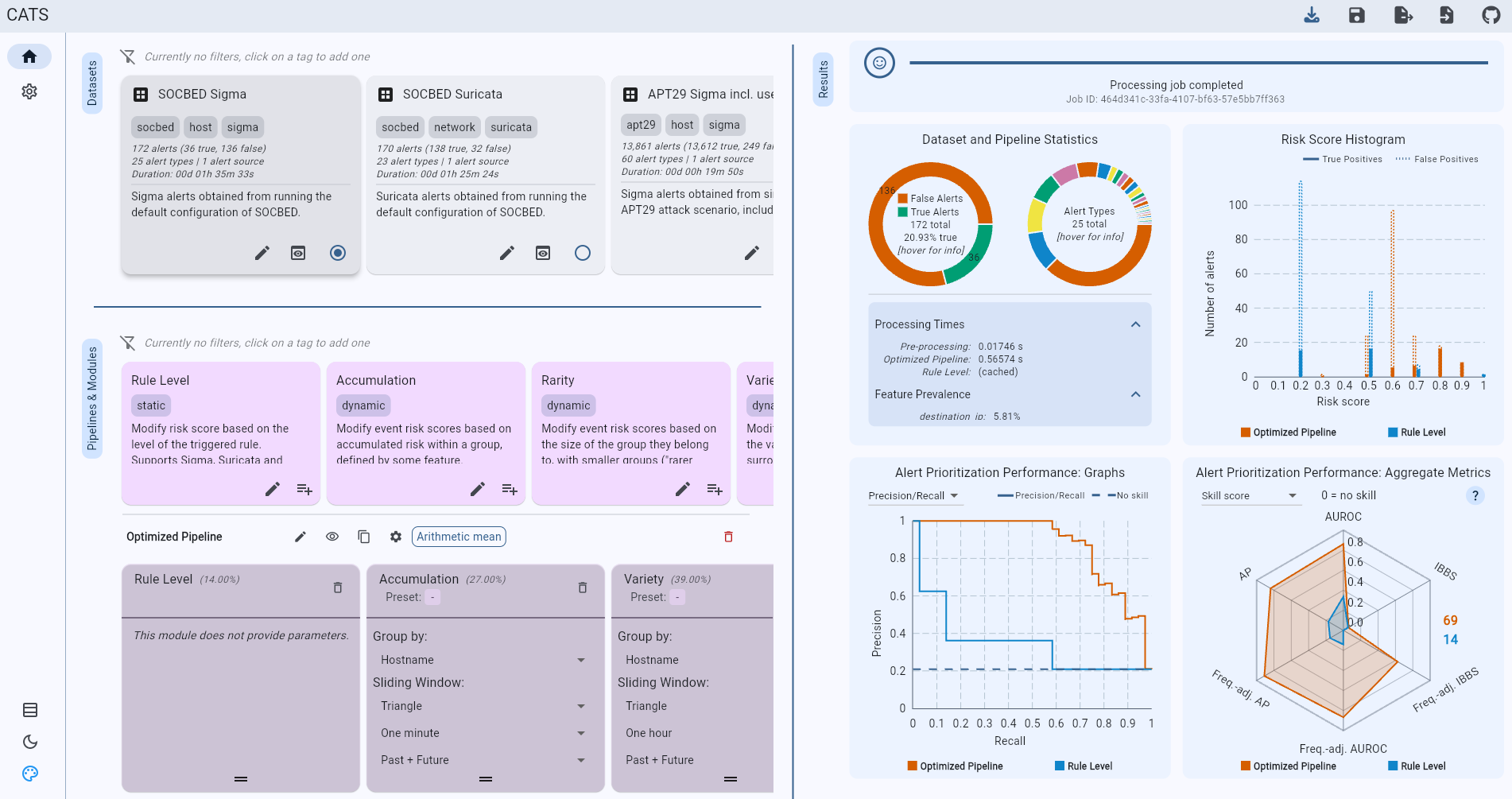}
    \caption{Screenshot of \name{}, our novel, free tool for evaluating and visually exploring cybersecurity alert prioritization methods such as risk-based alerting. The left side contains datasets, modules, and pipelines; the right side shows statistics and results.}
    \label{fig:cats}
\end{figure*}

\input{Alert}

\input{Heatmap2}

\section{CATS Web Interface}
\label{sec:catsscreenshot}

Figure~\ref{fig:cats} shows the main view of CATS' web UI.
Datasets can be added, viewed, and edited in the upper left area.
In particular, the raw JSON alerts as well as a jitter plot of all alerts, sorted by timestamp and type, can be inspected by clicking the ``Preview dataset'' button.
Below the datasets pane, all available risk modules are shown and can be edited.
At the bottom, pipelines can be added, configured, and populated with risk modules, which can also be configured in place.
For convenience, pipelines can be saved/loaded, exported/imported in JSON format, and duplicated.

The right side displays statistics and results.
Its upper left pane shows detailed dataset statistics, whereas the other three present the pipeline results: (1) a per-pipeline histogram of alert risk scores, divided by true and false alerts, (2) ROC, PR, F1, and MCC curves, including no-skill levels, and (3) skill-adjusted or unadjusted spider plots of the metrics described in Section~\ref{sec:metrics}.
All results can be saved in JSON format for further inspection or processing.

\section{CATS Alert Example}
\label{sec:catsalert}

CATS defines a straightforward format for alert datasets where each dataset is a human-readable JSON Lines file with one alert per line~\cite{githubcats}.
Each alert contains three top-level fields (\texttt{metadata}, \texttt{features}, and \texttt{full\_alert}), as shown in the example alert from the DEDALE Sigma dataset in Listing~\ref{lst:alert}.
The \texttt{full\_alert} field contains the original alert (here: a Sigma rule match), of which relevant fields are parsed by dedicated scripts to populate the \texttt{features} field.
The latter contains two mandatory (\texttt{timestamp}, \texttt{rule\_id}) and various optional fields (e.g., \texttt{hostname}), which are filled depending on their availability in the original alert.

\section{Discussion of Notable Risk Module Results}
\label{sec:notableresultsappendix}

This section discusses notable prioritization performances of individual risk modules as shown in Table~\ref{tab:heatmap}~(a).

The most striking pattern is the near-zero AUROC value of all Rarity and Aperiodicity instances for both AIT-ADS Wazuh and APT29S2 Sigma, indicating a \emph{reversed} prioritization.
Notably, these two datasets are the only ones containing neither Rarity nor Aperiodicity in their intra-dataset optimized pipeline (cf. Table~\ref{tab:stacked}).
The reason is similar for both:
AIT-ADS Wazuh is dominated by two alert types: ``Dovecot Authentication Success'' (all false, 62\% of total alerts) and ``Web server 400 error code'' (all true, 31\% of total alerts).
While the former alerts appear to be randomly distributed during the day, the latter occur in two short bursts within a two-minute attack window, which the fast Fourier transform underlying the Aperiodicity module rates as periodic activity.
Consequently, the benign alerts are both more aperiodic and more rare (with respect to the window length), resulting in the inverted prioritization.
Similarly, APT29S2 Sigma is dominated by two true alert types that make up 95\% of total alerts and are highly concentrated within a few bursts, leading to a neither aperiodic nor rare scoring.

These examples reveal a principal limitation of both the Rarity and Aperiodicity module, namely, their underlying assumption that true alerts are rare (which does not necessarily hold even if attacks are rare since one attack can cause thousands of alerts).
We thus recommend to not rely on these modules alone, but rather combine them with other modules (cf. Section~\ref{sec:rq2}) or dismiss them (cf. Appendix~\ref{sec:altpipeappendix}).

Relatedly, ERPCorp Falco stands out for the Rarity module with a large gap between the AUROC values for 1m (0.06) and 1h (0.69).
As discussed above, the reason appears to be a dense cluster of true alerts that are globally rare but not locally, which again indicates that the Rarity module should be used with longer windows to enable capturing deviations from longer-term normal activity.

Finally, another dataset that deserves closer inspection is APT29S2 Suricata.
It contains only two alert types, both having true and false instances, thus rendering the Rarity module ineffective.
This example shows that all datasets contribute to the evaluation's heterogeneity and that implementation decisions should not be based on single datasets.

\section{Pipeline Without Rarity and Aperiodicity}
\label{sec:altpipeappendix}

The mediocre prioritization performance of the Rarity and Aperiodicity modules (cf. Section~\ref{sec:rq1}) gives rise to the question whether they could be omitted from the pipeline.
We thus repeated the intra-dataset optimization and cross validation as described in Section~\ref{sec:rq2} but excluded these modules (i.e., only included Rule Level, Accumulation, and Variety).
The optimization yielded average weights of 12.5\% for Rule Level, 35\% for Accumulation, and 52.5\% for Variety, roughly preserving the ratio between these modules.

The results in Table~\ref{tab:heatmap2}~(a) show very similar results as the reference pipeline including the two modules, indicating that they might indeed be left out.
Notably, while the mean AUROC is subtly higher (0.93 instead of 0.92), AP is slightly lower (0.88 instead of 0.90).
Overall, the data is inconclusive as to which variant is better.
We thus advise practitioners to test both variants using their own alert data.

\section{Evaluation Without SOCBED and APT29S2}
\label{sec:altoptappendix}

Both the SOCBED- and APT29S2-based datasets are rather short (cf. Table~\ref{tab:datasets}).
In addition, APT29S2 Sigma has only 1.8\% false alerts and APT29S2 Suricata contains only two alert types, which might be unrealistic when compared to alerts in enterprise SOCs.
We therefore repeated the cross validation (cf. Section~\ref{sec:rq2}) but excluded these four datasets, yielding average weights of 15\% for Rule Level, 12.5\% for Accumulation, 50\% for Variety, 12.5\% for Rarity, and 10\% for Aperiodicity.
The weight changes are thus moderate as compared to the original evaluation (cf. Table~\ref{tab:stacked}), indicating that the pipeline is rather robust to dataset changes.

The resulting AUROC values are shown in Table~\ref{tab:heatmap2}~(b).
We can see that the individual results barely change for the remaining four datasets as compared to the reference pipeline.
Consequently, the mean AUROC is substantially higher because the SOCBED- and APT29S2-based datasets exhibited lower scores, showing that including these datasets in the original evaluation did not inflate the results, but instead even worsened them.
The fact that the optimized pipeline still performs quite well on these unusual datasets can be seen as an indication for its broad applicability.

\end{document}

%% file: RIRs.tex
\begin{table*}[!t]

    \caption{Literature on risk-based alerting proposes diverse rules for distinguishing true from false alerts. We generalize ten of them into three risk hypotheses (Accumulation, Variety, Rarity) and formulate two additional ones (Rule Level, Aperiodicity).}
    \label{tab:rirs}

    \footnotesize
    \setlength{\tabcolsep}{4pt}

    \newcommand{\rot}[1]{\makebox[11pt][l]{\rotatebox{45}{#1}}}
    \newcommand{\cm}{\checkmark}

    \begin{tabularx}{\textwidth}{c X c c c c c c c c c c c c c}

        \toprule

        \textbf{\#} & \textbf{Risk Incident Rule} &
            \rot{Apger~\cite{apger2018goodbye}} &
            \rot{Mills~\cite{mills2019building}} &
            \rot{Snyder~\cite{snyder2021proactive}} &
            \rot{Guide~\cite{splunk2024rba}} &
            \rot{GitHub~\cite{splunk2025githubrba}} &
            \rot{Docs~\cite{splunk2025how}} &
            &
            &
            \rot{\textbf{Accumulation}} &
            \rot{\textbf{Variety}} &
            \rot{\textbf{Rarity}} &
            \rot{\textbf{Other}} &
            \\

        \midrule

        1 & 24 Hour / 7 Day / 30 Day Risk Score Threshold Exceeded & \cm & \cm & \cm & \cm & & \cm & & & \cm & & & & \\ \grayrow
        2 & Anomalous Risk Score from Peers by Business Unit / Asset Category & & & \cm & \cm & & \cm & & & \cm & & & & \\
        3 & Increasing Risk for Role / Category / Threat Object / ATT\&CK Technique & & \cm & & \cm & & \cm & & & \cm & & & & \\ \grayrow
        4 & 24 Hour / 7 Day ATT\&CK Tactic / Technique Threshold Exceeded & \cm & \cm & \cm & \cm & & \cm & & & & \cm & & & \\
        5 & Entity with Risk from Multiple Source Types over 24 Hours / 7 Days / 30 Days & & \cm & & \cm & \cm & & & & & \cm & & & \\ \grayrow
        6 & Entity with Multiple Detections Within Single Source Type & & & & & \cm & & & & & \cm & & & \\
        7 & Multiple Risk Events Within Single ATT\&CK Tactic & \cm & & & & & & & & & \cm & & & \\ \grayrow
        8 & Threat Object Observed Across a Number of Risk Objects & & & & & & \cm & & & & \cm & & & \\
        9 & High Number of Unique ATT\&CK Techniques & & \cm & & & & & & & & \cm & & & \\ \grayrow
        10 & First Time Seen Threat Object Observed by Multiple Risk Objects (6 Months) & & & & \cm & & & & & & & \cm & & \\
        11 & Key Performance Indicator (KPI) Impact & & & & & & \cm & & & & & & \cm & \\ \grayrow
        12 & Mean Time to Resolution (MTTR) Threshold Exceeded & & & & & & \cm & & & & & & \cm & \\

        \bottomrule

    \end{tabularx}

\end{table*}

%% file: Modules.tex
\begin{table*}[!t]

    \caption{We implemented each of the five risk hypotheses as a parametrizable risk module in CATS. The parameters control which alerts are grouped together depending on their source, destination, or type, as well as their temporal proximity.}
    \label{tab:hypotheses}

    \newcommand{\cm}{\checkmark}
    \newcommand{\fs}{\ding{72}} %
    \newcommand{\na}{\textendash}

    \setlength{\tabcolsep}{5pt}
    \footnotesize

    \begin{tabularx}{\textwidth}{l X c c c c c c c c c c c c c c c c}

        \toprule

        & & \multicolumn{5}{c}{\textbf{Group By}}
        & & \multicolumn{10}{c}{\textbf{Window Length\,/\,Alignment\,/\,Shape}} \\
        \cmidrule{3-7} \cmidrule{9-18}
        \textbf{Risk Module}
        & \textbf{Underlying Risk Hypothesis}
        & H & U & S & D & A &
        & m & h & d & w &
        & c & t &
        & $\triangle$ & $\square$ \\

        \midrule

        Rule Level
        & Higher rule level implies higher risk
        & \na & \na & \na & \na & \na &
        & \na & \na & \na & \na &
        & \na & \na &
        & \na & \na \\ \grayrow
        Accumulation
        & Spatio-temporal accumulation of alerts implies higher risk
        & \fs & \cm & \fs & \cm & \cm &
        & \fs & \cm & \cm & \cm &
        & \fs & \cm &
        & \fs & \cm \\
        Variety
        & Spatio-temporal variety in alert types implies higher risk
        & \fs & \cm & \fs & \cm & \na &
        & \cm & \fs & \cm & \cm &
        & \fs & \cm &
        & \fs & \cm \\ \grayrow
        Rarity
        & Rarely occurring alerts imply higher risk
        & \cm & \cm & \cm & \cm & \fs &
        & \cm & \cm & \fs & \cm &
        & \fs & \cm &
        & \fs & \cm \\
        Aperiodicity
        & Aperiodically occurring alerts imply higher risk
        & \cm & \cm & \cm & \cm & \fs &
        & \cm & \cm & \fs & \cm &
        & \fs & \cm &
        & \na & \fs \\

        \bottomrule

    \end{tabularx}

    \vspace{10pt}
    \centering
    \cm\,=\,supported;\enspace
        \fs\,=\,default;\enspace
        \textbf{H}ostname;\enspace
        \textbf{U}sername;\enspace
        \textbf{S}rc\,IP;\enspace
        \textbf{D}st\,IP;\enspace
        \textbf{A}lert\,Type;\enspace
        \textbf{m}inute;\enspace
        \textbf{h}our;\enspace
        \textbf{d}ay;\enspace
        \textbf{w}eek;\enspace
        \textbf{c}entered;\enspace
        \textbf{t}railing;\enspace
        $\triangle$\,=\,triangle;\enspace
        $\square$\,=\,box

\end{table*}

%% file: Datasets.tex
\begin{table}[t]

    \caption{Our evaluation is based on eight alert datasets.}
    \label{tab:datasets}

    \footnotesize
    \setlength{\tabcolsep}{3.5pt}

    \begin{tabularx}{\columnwidth}{c X l r r r r c}

        \toprule

        \multicolumn{1}{l}{\makecell[l]{\textbf{\#}}} &
        \multicolumn{1}{l}{\makecell[l]{\textbf{Source}}} &
        \multicolumn{1}{l}{\makecell[c]{\textbf{Security}\\\textbf{System}}} &
        \multicolumn{1}{c}{\makecell[c]{\textbf{Alert}\\\textbf{Types}}} &
        \multicolumn{1}{c}{\makecell[c]{\textbf{Total}\\\textbf{Alerts}}} &
        \multicolumn{1}{c}{\makecell[c]{\textbf{True}\\\textbf{Alerts}}} &
        \multicolumn{1}{c}{\makecell[c]{\textbf{Base}\\\textbf{Rate}}} &
        \multicolumn{1}{c}{\makecell[c]{\textbf{Duration} $\blacktriangledown$\\\textbf{(dd:hh:mm)}}} \\

        \midrule

        1 & DEDALE & Sigma & 16 & 1\,819 & 51 & 3\% & 27:10:05 \\ \grayrow
        2 & \datasetname & Falco & 6 & 15\,060 & 1\,866 & 12\% & 05:19:09 \\
        3 & \multirow{2}{*}{AIT-ADS} & Wazuh & 20 & 23\,116 & 7\,661 & 33\% & 03:23:36 \\
        4 & & Suricata & 13 & 9\,186 & 44 & 1\% & 03:20:51 \\ \grayrow
        5 & & Sigma & 25 & 172 & 36 & 21\% & 00:01:35 \\ \grayrow
        6 & \multirow{-2}{*}{SOCBED} & Suricata & 23 & 170 & 138 & 81\% & 00:01:25 \\
        7 & \multirow{2}{*}{APT29S2} & Sigma & 60 & 13\,861 & 13\,612 & 98\% & 00:00:19 \\
        8 & & Suricata & 2 & 453 & 142 & 31\% & 00:00:11 \\

        \bottomrule

    \end{tabularx}

\end{table}

%% file: Heatmap.tex
\definecolor{hm000}{RGB}{255,  0,  0}
\definecolor{hm001}{RGB}{255,  5,  0}
\definecolor{hm002}{RGB}{255, 10,  0}
\definecolor{hm006}{RGB}{255, 31,  0}
\definecolor{hm007}{RGB}{255, 36,  0}
\definecolor{hm010}{RGB}{255, 51,  0}
\definecolor{hm017}{RGB}{255, 87,  0}
\definecolor{hm018}{RGB}{255, 92,  0}
\definecolor{hm021}{RGB}{255,107,  0}
\definecolor{hm023}{RGB}{255,117,  0}
\definecolor{hm026}{RGB}{255,133,  0}
\definecolor{hm029}{RGB}{255,148,  0}
\definecolor{hm031}{RGB}{255,158,  0}
\definecolor{hm033}{RGB}{255,168,  0}
\definecolor{hm035}{RGB}{255,179,  0}
\definecolor{hm036}{RGB}{255,184,  0}
\definecolor{hm037}{RGB}{255,189,  0}
\definecolor{hm038}{RGB}{255,194,  0}
\definecolor{hm040}{RGB}{255,204,  0}
\definecolor{hm041}{RGB}{255,209,  0}
\definecolor{hm042}{RGB}{255,214,  0}
\definecolor{hm045}{RGB}{255,230,  0}
\definecolor{hm047}{RGB}{255,240,  0}
\definecolor{hm048}{RGB}{255,245,  0}
\definecolor{hm049}{RGB}{255,250,  0}
\definecolor{hm050}{RGB}{255,255,  0}
\definecolor{hm053}{RGB}{241,248,  2}
\definecolor{hm054}{RGB}{237,246,  3}
\definecolor{hm055}{RGB}{233,244,  3}
\definecolor{hm056}{RGB}{228,242,  4}
\definecolor{hm057}{RGB}{224,240,  4}
\definecolor{hm058}{RGB}{220,238,  5}
\definecolor{hm059}{RGB}{215,236,  5}
\definecolor{hm062}{RGB}{202,230,  7}
\definecolor{hm063}{RGB}{198,228,  7}
\definecolor{hm064}{RGB}{193,226,  8}
\definecolor{hm066}{RGB}{184,222,  9}
\definecolor{hm067}{RGB}{180,220, 10}
\definecolor{hm068}{RGB}{175,218, 10}
\definecolor{hm069}{RGB}{171,216, 11}
\definecolor{hm071}{RGB}{162,212, 12}
\definecolor{hm072}{RGB}{158,210, 12}
\definecolor{hm073}{RGB}{153,208, 13}
\definecolor{hm074}{RGB}{149,206, 13}
\definecolor{hm075}{RGB}{145,197, 17}
\definecolor{hm076}{RGB}{140,195, 17}
\definecolor{hm079}{RGB}{127,189, 19}
\definecolor{hm080}{RGB}{123,187, 19}
\definecolor{hm081}{RGB}{118,185, 20}
\definecolor{hm084}{RGB}{105,179, 22}
\definecolor{hm085}{RGB}{100,177, 22}
\definecolor{hm086}{RGB}{ 96,175, 23}
\definecolor{hm087}{RGB}{ 91,173, 23}
\definecolor{hm088}{RGB}{ 87,171, 24}
\definecolor{hm091}{RGB}{ 74,165, 26}
\definecolor{hm092}{RGB}{ 70,163, 26}
\definecolor{hm094}{RGB}{ 61,159, 27}
\definecolor{hm095}{RGB}{ 57,157, 28}
\definecolor{hm096}{RGB}{ 52,155, 28}
\definecolor{hm097}{RGB}{ 48,153, 29}
\definecolor{hm099}{RGB}{ 39,149, 30}
\definecolor{hm100}{RGB}{ 34,139, 34}

\newcommand{\cc}[2]{\cellcolor{#1}#2}
\newcommand{\gc}[1]{\cellcolor{gray!10}#1}

\newlength{\ccw}
\setlength{\ccw}{1.19cm}
\newcolumntype{H}{>{\centering\arraybackslash}m{\ccw}}

\newcommand{\tikzbrace}[2]{%
    \multirow{#1}{*}{%
        \begin{tikzpicture}[baseline=(label.center)]
            \node[right] at (0,{#1*\baselineskip*\arraystretch*0.5+#1*\extrarowheight*0.5}) (label) {\textbf{#2}};
            \draw[decorate,decoration={calligraphic brace,amplitude=4pt},line width=1.0pt]
            (2.4em,0) -- (2.4em,{#1*\baselineskip*\arraystretch+#1*\extrarowheight});
        \end{tikzpicture}%
    }%
}

\begin{table*}[!t]

    \caption{Alert prioritization performance of the evaluated risk modules and their combinations (pipelines) across datasets, with different window lengths and otherwise default parametrization (AUROC, 1\,=\,perfect prioritization, 0.5\,=\,chance, 0\,=\,inverted).}
    \label{tab:heatmap}

    \setlength{\tabcolsep}{0pt}
    \footnotesize

    \begin{tabularx}{\textwidth}{m{0.75cm} X l H H H H H H H H  c H H H}

        \toprule

        \multicolumn{2}{l}{\textbf{Alert Prioritization Method}} & &
        \rotatebox{45}{\shortstack[l]{DEDALE\\Sigma}} &
        \rotatebox{45}{\shortstack[l]{ERPCorp\\Falco}} &
        \rotatebox{45}{\shortstack[l]{AIT-ADS\\Wazuh}} &
        \rotatebox{45}{\shortstack[l]{AIT-ADS\\Suricata}} &
        \rotatebox{45}{\shortstack[l]{SOCBED\\Sigma}} &
        \rotatebox{45}{\shortstack[l]{SOCBED\\Suricata}} &
        \rotatebox{45}{\shortstack[l]{APT29S2\\Sigma}} &
        \rotatebox{45}{\shortstack[l]{APT29S2\\Suricata}} &
        \hspace{10pt} &
        \rotatebox{45}{\textbf{Mean}} &
        \rotatebox{45}{\textbf{Std}} &
        \rotatebox{45}{\textbf{\textgreater\,0.5}} \\

        \midrule

        & No Prioritization & &
        \cc{hm050}{0.50} & \cc{hm050}{0.50} & \cc{hm050}{0.50} & \cc{hm050}{0.50} &
        \cc{hm050}{0.50} & \cc{hm050}{0.50} & \cc{hm050}{0.50} & \cc{hm050}{0.50} & &
        \cc{hm050}{0.50} & \gc{0.00} & \gc{0\,/\,8} \\

        \midrule

        \tikzbrace{14}{(a)} &
        Rule Level & &
        \cc{hm053}{0.53} & \cc{hm050}{0.50} & \cc{hm099}{0.99} & \cc{hm081}{0.81} &
        \cc{hm067}{0.67} & \cc{hm100}{1.00} & \cc{hm074}{0.74} & \cc{hm050}{0.50} & &
        \cc{hm072}{0.72} & \gc{0.21} & \gc{6\,/\,8} \\

        \addlinespace[3pt]

        & \multirow{3}{*}{Accumulation}
        & 1m\hspace{10pt} &
        \cc{hm084}{0.84} & \cc{hm099}{0.99} & \cc{hm100}{1.00} & \cc{hm066}{0.66} &
        \cc{hm033}{0.33} & \cc{hm091}{0.91} & \cc{hm099}{0.99} & \cc{hm079}{0.79} & &
        \cc{hm081}{0.81} & \gc{0.23} & \gc{7\,/\,8} \\
        & & 1h &
        \cc{hm085}{0.85} & \cc{hm097}{0.97} & \cc{hm100}{1.00} & \cc{hm048}{0.48} &
        \cc{hm056}{0.56} & \cc{hm095}{0.95} & \cc{hm099}{0.99} & \cc{hm000}{0.00} & &
        \cc{hm072}{0.72} & \gc{0.36} & \gc{6\,/\,8} \\
        & & 1d &
        \cc{hm086}{0.86} & \cc{hm100}{1.00} & \cc{hm100}{1.00} & \cc{hm055}{0.55} &
        \cc{hm062}{0.62} & \cc{hm095}{0.95} & \cc{hm099}{0.99} & \cc{hm000}{0.00} & &
        \cc{hm075}{0.75} & \gc{0.35} & \gc{7\,/\,8} \\

        \addlinespace[3pt]

        & \multirow{3}{*}{Variety}
        & 1m &
        \cc{hm049}{0.49} & \cc{hm054}{0.54} & \cc{hm100}{1.00} & \cc{hm080}{0.80} &
        \cc{hm064}{0.64} & \cc{hm067}{0.67} & \cc{hm094}{0.94} & \cc{hm079}{0.79} & &
        \cc{hm073}{0.73} & \gc{0.18} & \gc{7\,/\,8} \\
        & & 1h &
        \cc{hm092}{0.92} & \cc{hm100}{1.00} & \cc{hm100}{1.00} & \cc{hm096}{0.96} &
        \cc{hm096}{0.96} & \cc{hm069}{0.69} & \cc{hm087}{0.87} & \cc{hm053}{0.53} & &
        \cc{hm087}{0.87} & \gc{0.17} & \gc{8\,/\,8} \\
        & & 1d &
        \cc{hm099}{0.99} & \cc{hm100}{1.00} & \cc{hm100}{1.00} & \cc{hm079}{0.79} &
        \cc{hm094}{0.94} & \cc{hm069}{0.69} & \cc{hm088}{0.88} & \cc{hm053}{0.53} & &
        \cc{hm085}{0.85} & \gc{0.17} & \gc{8\,/\,8} \\

        \addlinespace[3pt]

        & \multirow{3}{*}{Rarity}
        & 1m &
        \cc{hm029}{0.29} & \cc{hm006}{0.06} & \cc{hm000}{0.00} & \cc{hm059}{0.59} &
        \cc{hm092}{0.92} & \cc{hm056}{0.56} & \cc{hm001}{0.01} & \cc{hm001}{0.01} & &
        \cc{hm031}{0.31} & \gc{0.35} & \gc{3\,/\,8} \\
        & & 1h &
        \cc{hm033}{0.33} & \cc{hm069}{0.69} & \cc{hm000}{0.00} & \cc{hm088}{0.88} &
        \cc{hm096}{0.96} & \cc{hm068}{0.68} & \cc{hm001}{0.01} & \cc{hm002}{0.02} & &
        \cc{hm045}{0.45} & \gc{0.41} & \gc{4\,/\,8} \\
        & & 1d &
        \cc{hm040}{0.40} & \cc{hm072}{0.72} & \cc{hm007}{0.07} & \cc{hm099}{0.99} &
        \cc{hm092}{0.92} & \cc{hm045}{0.45} & \cc{hm002}{0.02} & \cc{hm026}{0.26} & &
        \cc{hm048}{0.48} & \gc{0.37} & \gc{3\,/\,8} \\

        \addlinespace[3pt]

        & \multirow{3}{*}{Aperiodicity}
        & 1m &
        \cc{hm054}{0.54} & \cc{hm066}{0.66} & \cc{hm010}{0.10} & \cc{hm042}{0.42} &
        \cc{hm058}{0.58} & \cc{hm076}{0.76} & \cc{hm006}{0.06} & \cc{hm066}{0.66} & &
        \cc{hm047}{0.47} & \gc{0.26} & \gc{5\,/\,8} \\
        & & 1h &
        \cc{hm074}{0.74} & \cc{hm071}{0.71} & \cc{hm000}{0.00} & \cc{hm072}{0.72} &
        \cc{hm092}{0.92} & \cc{hm076}{0.76} & \cc{hm002}{0.02} & \cc{hm050}{0.50} & &
        \cc{hm055}{0.55} & \gc{0.35} & \gc{5\,/\,8} \\
        & & 1d &
        \cc{hm075}{0.75} & \cc{hm072}{0.72} & \cc{hm001}{0.01} & \cc{hm095}{0.95} &
        \cc{hm085}{0.85} & \cc{hm076}{0.76} & \cc{hm002}{0.02} & \cc{hm050}{0.50} & &
        \cc{hm057}{0.57} & \gc{0.37} & \gc{5\,/\,8} \\

        \midrule

        \tikzbrace{2}{(b)}
        & \multicolumn{2}{l}{Opt. Pipeline (intra-dataset)} &
        \cc{hm100}{1.00} & \cc{hm100}{1.00} & \cc{hm100}{1.00} & \cc{hm100}{1.00} &
        \cc{hm097}{0.98} & \cc{hm100}{1.00} & \cc{hm099}{0.99} & \cc{hm079}{0.79} & &
        \cc{hm097}{0.97} & \gc{0.07} & \gc{8\,/\,8} \\
        & \multicolumn{2}{l}{Opt. Pipeline (leave-one-out)} &
        \cc{hm100}{1.00} & \cc{hm100}{1.00} & \cc{hm097}{0.98} & \cc{hm099}{0.99} &
        \cc{hm092}{0.93} & \cc{hm091}{0.90} & \cc{hm084}{0.84} & \cc{hm073}{0.73} & &
        \cc{hm092}{0.92} & \gc{0.09} & \gc{8\,/\,8} \\

        \midrule

        \tikzbrace{3}{(c)}
        & Geometric Mean & &
        \cc{hm100}{1.00} & \cc{hm100}{1.00} & \cc{hm085}{0.85} & \cc{hm097}{0.98} &
        \cc{hm091}{0.90} & \cc{hm088}{0.89} & \cc{hm094}{0.94} & \cc{hm069}{0.70} & &
        \cc{hm091}{0.91} & \gc{0.10} & \gc{8\,/\,8} \\
        & Box Window & &
        \cc{hm097}{0.97} & \cc{hm100}{1.00} & \cc{hm087}{0.87} & \cc{hm100}{1.00} &
        \cc{hm097}{0.97} & \cc{hm091}{0.90} & \cc{hm057}{0.57} & \cc{hm067}{0.67} & &
        \cc{hm087}{0.87} & \gc{0.16} & \gc{8\,/\,8} \\
        & Trailing Window & &
        \cc{hm059}{0.60} & \cc{hm100}{1.00} & \cc{hm099}{0.99} & \cc{hm097}{0.97} &
        \cc{hm053}{0.53} & \cc{hm075}{0.75} & \cc{hm068}{0.68} & \cc{hm048}{0.48} & &
        \cc{hm075}{0.75} & \gc{0.21} & \gc{7\,/\,8} \\

        \bottomrule

    \end{tabularx}

\end{table*}

%% file: Stacked.tex
\begin{table}[t]

    \caption{Risk module weights of the optimized pipelines. The optimization shows that all five modules contribute positively to the prioritization performance for at least some datasets.}
    \label{tab:stacked}

    \definecolor{catA}{HTML}{4878A8}
    \definecolor{catB}{HTML}{5DA5A8}
    \definecolor{catC}{HTML}{7AB648}
    \definecolor{catD}{HTML}{E8A838}
    \definecolor{catE}{HTML}{D06040}

    \newlength{\barwidth}
    \newlength{\segwidth}
    \newcommand{\slbar}[3]{%
        \setlength{\segwidth}{#2\barwidth}%
        \makebox[0pt][l]{\raisebox{-1.5pt}{\textcolor{#1}{\rule{\segwidth}{9pt}}}}%
        \makebox[\segwidth][c]{\bfseries\textcolor{white}{#3}}%
    }
    \setlength{\barwidth}{0.33\columnwidth}

    \footnotesize
    \setlength{\tabcolsep}{0pt}

    \begin{tabularx}{\columnwidth}{X l l l}

        \toprule

        \textbf{Dataset} & \textbf{Intra-Dataset Opt.} & \hspace{10pt} & \textbf{Leave-One-Out Mean} \\

        \midrule

        DEDALE Sigma &
            \slbar{catB}{0.30}{30}%
            \slbar{catC}{0.30}{30}%
            \slbar{catE}{0.20}{20}%
            \slbar{catD}{0.20}{20} & &
            \slbar{catA}{0.14}{14}%
            \slbar{catB}{0.23}{23}%
            \slbar{catC}{0.46}{46}%
            \slbar{catE}{0.08}{8}%
            \slbar{catD}{0.09}{9} \\
        ERPCorp Falco &
            \slbar{catC}{0.70}{70}%
            \slbar{catE}{0.10}{10}%
            \slbar{catD}{0.20}{20} & &
            \slbar{catA}{0.14}{14}%
            \slbar{catB}{0.27}{27}%
            \slbar{catC}{0.40}{40}%
            \slbar{catE}{0.10}{10}%
            \slbar{catD}{0.09}{9} \\
        AIT-ADS Wazuh &
            \slbar{catA}{0.40}{40}%
            \slbar{catB}{0.20}{20}%
            \slbar{catC}{0.40}{40} & &
            \slbar{catA}{0.09}{9}%
            \slbar{catB}{0.24}{24}%
            \slbar{catC}{0.44}{44}%
            \slbar{catE}{0.11}{11}%
            \slbar{catD}{0.12}{12} \\
        AIT-ADS Suricata &
            \slbar{catA}{0.20}{20}%
            \slbar{catC}{0.60}{60}%
            \slbar{catE}{0.20}{20} & &
            \slbar{catA}{0.12}{12}%
            \slbar{catB}{0.27}{27}%
            \slbar{catC}{0.41}{41}%
            \slbar{catE}{0.09}{9}%
            \slbar{catD}{0.11}{11} \\
        SOCBED Sigma &
            \slbar{catC}{0.80}{80}%
            \slbar{catE}{0.20}{20} & &
            \slbar{catA}{0.14}{14}%
            \slbar{catB}{0.27}{27}%
            \slbar{catC}{0.39}{39}%
            \slbar{catE}{0.09}{9}%
            \slbar{catD}{0.11}{11} \\
        SOCBED Suricata &
            \slbar{catA}{0.30}{30}%
            \slbar{catB}{0.10}{10}%
            \slbar{catC}{0.30}{30}%
            \slbar{catD}{0.30}{30} & &
            \slbar{catA}{0.10}{10}%
            \slbar{catB}{0.26}{26}%
            \slbar{catC}{0.46}{46}%
            \slbar{catE}{0.11}{11}%
            \slbar{catD}{0.07}{7} \\
        APT29S2 Sigma &
            \slbar{catA}{0.10}{10}%
            \slbar{catB}{0.90}{90} & &
            \slbar{catA}{0.13}{13}%
            \slbar{catB}{0.14}{14}%
            \slbar{catC}{0.50}{50}%
            \slbar{catE}{0.11}{11}%
            \slbar{catD}{0.12}{12} \\
        APT29S2 Suricata &
            \slbar{catB}{0.40}{40}%
            \slbar{catC}{0.40}{40}%
            \slbar{catE}{0.10}{10}%
            \slbar{catD}{0.10}{10} & &
            \slbar{catA}{0.14}{14}%
            \slbar{catB}{0.22}{22}%
            \slbar{catC}{0.44}{44}%
            \slbar{catE}{0.10}{10}%
            \slbar{catD}{0.10}{10} \\

        \midrule

        \textbf{Mean} &
            \slbar{catA}{0.12}{12}%
            \slbar{catB}{0.24}{24}%
            \slbar{catC}{0.44}{44}%
            \slbar{catE}{0.10}{10}%
            \slbar{catD}{0.10}{10} & &
            \slbar{catA}{0.12}{12}%
            \slbar{catB}{0.24}{24}%
            \slbar{catC}{0.44}{44}%
            \slbar{catE}{0.10}{10}%
            \slbar{catD}{0.10}{10} \\

        \bottomrule

    \end{tabularx}

    \centering
    \vspace{10pt}
    \textcolor{catA}{\rule{6pt}{6pt}}~\textsf{Rule\,Level}\enspace
        \textcolor{catB}{\rule{6pt}{6pt}}~\textsf{Accum.\,1m}\enspace
        \textcolor{catC}{\rule{6pt}{6pt}}~\textsf{Variety\,1h}\enspace
        \textcolor{catE}{\rule{6pt}{6pt}}~\textsf{Rarity\,1d}\enspace
        \textcolor{catD}{\rule{6pt}{6pt}}~\textsf{Aperiod.\,1d}

\end{table}

%% file: Alert.tex
\lstdefinestyle{json}{
    basicstyle=\ttfamily\scriptsize,
    frame=single,
    breaklines=true,
    columns=fullflexible,
    showstringspaces=false,
    captionpos=b,
    abovecaptionskip=0.5\baselineskip,
}

\lstset{
    string=[s]{"}{"},
    stringstyle=\color{violet},
    comment=[l]{:},
    commentstyle=\color{black},
}

\begin{figure*}[p]
    \begin{lstlisting}[
        style=json,
        caption={Example alert in \name{} JSON format. The \texttt{features} field contains information parsed from the original Sigma alert (\texttt{full\_alert} field). \texttt{metadata} defines alert source, serial number, and unique ID as well as the misuse label (\texttt{true} or \texttt{false}).},
        label={lst:alert}
        ]
{
    "metadata": {
        "alert_source": "dedale_sigma",
        "alert_index": 295,
        "alert_id": "520e625b-ca13-4174-b30a-9924e9be35b6",
        "misuse": true
    },
    "features": {
        "timestamp": "2025-01-06T11:00:05.905Z",
        "hostname": "CLIENT2",
        "username": "client2",
        "rule_name": "Potential Persistence Attempt Via Run Keys Using Reg.EXE",
        "rule_id": "de587dce-915e-4218-aac4-835ca6af6f70",
        "rule_level": "medium",
        "rule_attacktechniques": ["t1547"],
        "rule_attacktactics": ["privilege","persistence"]
    },
    "full_alert": {
        "rule_title": "Potential Persistence Attempt Via Run Keys Using Reg.EXE",
        "rule_id": "de587dce-915e-4218-aac4-835ca6af6f70",
        "rule_level": "medium",
        "tags": ["attack.privilege-escalation","attack.persistence","attack.t1547.001"],
        "log": {
            "Provider_Name": "Microsoft-Windows-Sysmon",
            "EventID": 1,
            "EventType": "Process Create (rule: ProcessCreate)",
            "Image": "C:\\Windows\\System32\\reg.exe",
            "CommandLine": "reg  add \"HKCU\\Software\\Microsoft\\Windows\\CurrentVersion\\Run\" /v WSQtxiUDesrgQJ /t REG_SZ /d C:\\Users\\client2\\AppData\\Local\\Temp\\svcmon.exe /f",
            ...
        }
    }
}
    \end{lstlisting}
\end{figure*}

%% file: Heatmap2.tex
\setlength{\ccw}{1.16cm}

\begin{table*}[!t]

    \caption{AUROC scores for two alternative optimized leave-one-out pipelines, showing that (a)~the Rarity and Aperiodicity modules might be dispensable and (b)~the rather short SOCBED- and APT29S2-based datasets do not inflate evaluation results.}
    \label{tab:heatmap2}

    \newcommand{\na}{\textendash}

    \setlength{\tabcolsep}{0pt}
    \footnotesize

    \begin{tabularx}{\textwidth}{m{0.475cm} X H H H H H H H H  c H H H}

        \toprule

        \multicolumn{2}{l}{\textbf{Alert Prioritization Method}} &
        \rotatebox{45}{\shortstack[l]{DEDALE\\Sigma}} &
        \rotatebox{45}{\shortstack[l]{ERPCorp\\Falco}} &
        \rotatebox{45}{\shortstack[l]{AIT-ADS\\Wazuh}} &
        \rotatebox{45}{\shortstack[l]{AIT-ADS\\Suricata}} &
        \rotatebox{45}{\shortstack[l]{SOCBED\\Sigma}} &
        \rotatebox{45}{\shortstack[l]{SOCBED\\Suricata}} &
        \rotatebox{45}{\shortstack[l]{APT29S2\\Sigma}} &
        \rotatebox{45}{\shortstack[l]{APT29S2\\Suricata}} &
        \hspace{10pt} &
        \rotatebox{45}{\textbf{Mean}} &
        \rotatebox{45}{\textbf{Std}} &
        \rotatebox{45}{\textbf{\textgreater\,0.5}} \\

        \midrule

        & Reference Pipeline (cf. Table~\ref{tab:heatmap}) &
        \cc{hm100}{1.00} & \cc{hm100}{1.00} & \cc{hm097}{0.98} & \cc{hm099}{0.99} &
        \cc{hm092}{0.93} & \cc{hm091}{0.90} & \cc{hm084}{0.84} & \cc{hm073}{0.73} & &
        \cc{hm092}{0.92} & \gc{0.09} & \gc{8\,/\,8} \\

        \textbf{(a)} &
        Without Rarity and Aperiodicity &
        \cc{hm096}{0.96} & \cc{hm100}{1.00} & \cc{hm100}{1.00} & \cc{hm095}{0.95} &
        \cc{hm081}{0.82} & \cc{hm096}{0.96} & \cc{hm099}{0.99} & \cc{hm079}{0.79} & &
        \cc{hm092}{0.93} & \gc{0.08} & \gc{8\,/\,8} \\

        \textbf{(b)} &
        Without SOCBED and APT29S2 &
        \cc{hm099}{0.99} & \cc{hm100}{1.00} & \cc{hm097}{0.97} & \cc{hm100}{1.00} &
        \gc{\na} & \gc{\na} & \gc{\na} & \gc{\na} & &
        \cc{hm099}{0.99} & \gc{0.01} & \gc{4\,/\,4} \\

        \bottomrule

    \end{tabularx}

\end{table*}